\documentclass[a4paper,11pt]{article}
\usepackage{jheppub} 
\usepackage{lineno}
\usepackage[dvipsnames]{xcolor}

\usepackage{caption}
\usepackage{subcaption}
\usepackage{mathtools}
\DeclarePairedDelimiter\abs{\lvert}{\rvert}
\usepackage{slashed}
\usepackage{cleveref}
\usepackage[normalem]{ulem}
\usepackage{multirow}

\usepackage{tablefootnote} 

\arxivnumber{2405.08081} 

\title{\boldmath Not-so-inelastic Dark Matter}

\preprint{P3H-24-028, TTP24-011}

\author[1]{Giovani~Dalla~Valle~Garcia}
\author[2]{Felix~Kahlhoefer}
\author[1,3]{Maksym~Ovchynnikov}
\author[1]{Thomas~Schwetz}
\affiliation[1]{Institut für Astroteilchen Physik, Karlsruher Institut für Technologie (KIT), Hermann-von-\\Helmholtz-Platz 1, 76344 Eggenstein-Leopoldshafen, Germany}
\affiliation[2]{Institute for Theoretical Particle Physics (TTP), Karlsruhe Institute of Technology (KIT), \\D-76131 Karlsruhe, Germany}
\affiliation[3]{Instituut-Lorentz, Leiden University, Niels Bohrweg 2, 2333 CA Leiden, The Netherlands}

\emailAdd{giovani.garcia@student.kit.edu}
\emailAdd{kahlhoefer@kit.edu}
\emailAdd{maksym.ovchynnikov@kit.edu}
\emailAdd{schwetz@kit.edu}

{\abstract{Models of inelastic (or pseudo-Dirac) dark matter commonly assume an accidental symmetry between the left-handed and right-handed mass terms in order to suppress diagonal couplings. We point out that this symmetry is unnecessary because for Majorana fermions the diagonal couplings are not strongly constrained. Removing the requirement of such an ad hoc symmetry instead relaxes the relic density constraint due to additional annihilation modes. We consider a simple UV-complete model realizing this setup and study constraints from (in)direct detection, beam dump experiments and colliders. We identify two viable mass regions for the dark matter mass, around a few hundred MeV and around a few GeV, respectively. The former region will be fully tested by near-future analyses of the NA64 and Belle II data, while the latter turns out to be challenging to explore even with future experiments. 
}

\keywords{ Beyond the Standard Model: Dark Matter at Colliders, Models for Dark Matter, New Light Particles, Cosmology of Theories BSM }

}

\begin{document}
\maketitle
\flushbottom

{\section{Introduction}
\label{sec:intro}
The idea that Dark Matter (DM) particles can only interact with Standard Model (SM) particles through an inelastic transition from a lighter to a heavier (excited) state was first proposed as a possible interpretation of the DAMA annual modulation signal~\cite{DAMA:2000mdu} consistent with null results from other direct detection experiments~\cite{Tucker-Smith:2001myb,Tucker-Smith:2004mxa,Chang:2008gd,Schmidt-Hoberg:2009sgp,Schwetz:2011xm,Bozorgnia:2013hsa}. In these early models of inelastic DM (iDM), the mass splitting was assumed to be comparable to the kinetic energy of DM particles bound to the Milky Way, such that upscattering is suppressed but not impossible, leading to a modified recoil spectrum and an attractive target for direct detection experiments~\cite{Bernabei:2002qr,XENON10:2009sho,CDMS-II:2010wvq,XENON100:2011hxw,PandaX-II:2017zex,XENON:2020fgj}. But as attention shifted away from the DAMA anomaly, it was soon realized that iDM has intriguing implications for many other situations, such as stellar evolution~\cite{Hooper:2010es,McCullough:2010ai,Bell:2018pkk} or structure formation~\cite{Blennow:2016gde,Vogelsberger:2018bok,Alvarez:2019nwt,Essig:2018pzq,ONeil:2022szc}, and that one can also search for the de-excitation or downscattering of excited states~\cite{Finkbeiner:2007kk,Feldstein:2010su,Finkbeiner:2014sja,Eby:2019mgs,Baryakhtar:2020rwy,Emken:2021vmf,DEramo:2016gqz}.

Interest in iDM models has rapidly grown in recent years due to a combination of two of their main properties. First of all, excited states are typically short-lived on cosmological timescales, such that DM annihilation is strongly suppressed and constraints from the Cosmic Microwave Background~\cite{Planck:2018vyg} and searches for x-ray emission~\cite{Cirelli:2023tnx} are evaded. This makes it possible to consider thermally produced DM particles in the sub-GeV mass range, which are otherwise in strong tension with these constraints~\cite{Battaglieri:2017aum,Beacham:2019nyx}.\footnote{An alternative avenue is to explore non-thermal production of iDM via the freeze-in mechanism~\cite{An:2020tcg,Heeba:2023bik}. In this case the annihilation cross section is extremely small and the excited state may remain populated on cosmological scales, see also ref.~\cite{CarrilloGonzalez:2021lxm,Brahma:2023psr}.} Second, excited states can be long-lived on laboratory timescales, such that the production of excited states in accelerator experiments can give rise to striking signatures, such as displaced vertices in the detector~\cite{Davoli:2017swj,Izaguirre:2015zva,Tsai:2019buq,Duerr:2019dmv,Kang:2021oes,Davighi:2024zip}. Models of iDM therefore open up new parameter regions that can be explored with a broad range of experimental strategies~\cite{Berlin:2018jbm,Berlin:2018bsc,Batell:2021ooj,Bertuzzo:2022ozu,Mongillo:2023hbs}.

One of the simplest realizations of iDM requires two chiral fermions $\chi_L$ and $\chi_R$ with Majorana mass terms $m_L$ and $m_R$ (typically generated through a dark Higgs mechanism~\cite{Duerr:2020muu}) and a joint Dirac mass term $m_d$~\cite{DeSimone:2010tf,Duerr:2019dmv}.\footnote{More complex models are discussed for example in refs.~\cite{Filimonova:2022pkj,Abdullahi:2023tyk}.}  For $m_d \gg m_L, m_R$ one obtains a so-called Pseudo-Dirac fermion, i.e.\ two mass eigenstates with small mass splitting. The ``inelasticity'' of the model, i.e.\ the absence of interactions involving only the ground state, results from assuming $m_L \approx m_R$. This symmetry is however difficult to motivate for DM particles that couple to SM fermions, which do not have such a symmetry~\cite{Abdullahi:2023tyk}.

In this work we point out that the requirement $m_L \approx m_R$ is in fact unnecessary. This is because the presence of elastic (diagonal) interactions does not actually spoil any of the attractive features of iDM models. Since the two mass eigenstates are Majorana particles, their scattering and annihilation cross sections are strongly suppressed in the non-relativistic limit, such that the diagonal couplings are not strongly constrained by experiments. In the early universe, on the other hand, the presence of diagonal couplings significantly enhances the annihilation rate of relativistic DM particles, and thereby relaxes the constraints on the model from the requirement that the observed DM relic abundance must be reproduced.

We propose a new model, called not-so-inelastic DM (niDM), which has one new parameter, called $\delta_y$, characterising the difference between $m_L$ and $m_R$, such that the iDM case is recovered in the limit $\delta_y \to 0$. For $\delta_y \gg 1$, however, the model sustantially differs from iDM, because the relaxed relic density requirement means that much larger values of the mass splitting can be compatible with the relic density requirement. In fact, the niDM model interpolates between iDM models and see-saw-like models, where the mass splitting is so large that the heavier state becomes irrelevant for phenomenology and a single Majorana DM (mDM) particle remains~\cite{Battaglieri:2017aum}. The niDM model is therefore a straight-forward extension of the model discussed in ref.~\cite{Kahlhoefer:2015bea,Duerr:2016tmh} with a second fermion in the dark sector as required by anomaly cancellation to make the model self-consistent and UV complete.

In our analysis we consider specifically the case where the DM particles interact with SM fermions through a dark photon arising from a spontaneously broken $U(1)'$ symmetry with kinetic mixing with the SM photon~\cite{Holdom:1985ag}. 
Dark photon models face tight constraints from missing energy searches at electron-positron colliders. In our set-up such signatures arise when the dark photon decays into a pair of lighter states via the diagonal coupling, or if the excited state produced via the off-diagonal coupling decays outside of the detector acceptance. We calculate the resulting constraints from BaBar~\cite{BaBar:2017tiz} and estimate the sensitivity of Belle II~\cite{Belle-II:2018jsg}. We also discuss the relevance of ongoing beam dump experiments such as NA62~\cite{NA62:2023qyn} and NA64~\cite{NA64:2023ehh}, and analyse (in)direct detection bounds due to the now present elastic (diagonal) interactions.

The remainder of this work is structured as follows. In section~\ref{sec:modeltext} we introduce the model that we study in this work, derive relevant relations between different model parameters and identify the parameter regions of particular phenomenological interest. We discuss the decay modes of the heavier fermion in section~\ref{sec:decays}. Section~\ref{sec:relic} is then devoted to a detailed discussion of the relic density calculation in this model, while section~\ref{sec:detection} considers the sensitivity of direct and indirect detection experiments. In section~\ref{sec:colliders} we discuss the sensitivity of collider and accelerator experiments. We present our main results in section~\ref{sec:results} before concluding in section~\ref{sec:conclu}. In appendix~\ref{app:general-model} we discuss the niDM model in full generality, whereas in appendices~\ref{app:beam-dump}, \ref{app:NA64details}, \ref{app:colliders} we provide details on our analyses of beam dump experiments, the NA64 recast, and electron-positron colliders, respectively.
}

{
\section{A simple model of not-so-inelastic Dark Matter (niDM)}
\label{sec:modeltext}

In this section we introduce a simple extension of the SM in order to implement niDM and discuss the parameters relevant for its phenomenology. The model can be thought of as either a generalisation of the coupling structure of models of inelastic DM (iDM) or as an extension of the two-mediator Majorana DM (mDM) model considered in~ refs.~\cite{Kahlhoefer:2015bea,Duerr:2016tmh}, where a second 2-component fermion is introduced for anomaly cancellation.

We consider a dark sector containing a new $U(1)'$ gauge symmetry. The dark sector contains two 2-component fermion fields of opposite chirality, $\chi_L$ and $\chi_R$, which are singlets under the SM gauge group but carry the same $U(1)'$ charge $q_{\chi}$ such that our extension of the SM is anomaly-free~\cite{Costa:2020krs}. Furthermore, we introduce a SM-singlet scalar field $S$ with charge $q_s = - 2 q_{\chi}$ under  $U(1)'$ which acquires a vacuum expectation value (vev) $\langle S\rangle = w/\sqrt{2}$. The vev breaks the new gauge symmetry spontaneously and generates a mass for the corresponding $A'$ gauge boson, as well as Majorana mass terms for both $\chi_L$ and $\chi_R$. The most general Lagrangian for the new fields together with the SM fields allowed by the gauge symmetries can be written as 
\begin{equation} \label{eq:NewPhysicsLagrangian}
    \mathcal{L}_{\text{NP}} =  \mathcal{L}_{\chi} + \mathcal{L}_{V} + \mathcal{L}_{S}  \, ,
\end{equation} 
where
\begin{align} 
    \mathcal{L}_{\chi}  = & i \bar{\chi}_L \slashed{D} \chi_L + i \bar{\chi}_R \slashed{D} \chi_R - m_D^{\ast} \bar\chi_L\chi_R - \sqrt{2} y_L S \bar{\chi}^c_L \chi_L - \sqrt{2} y_R S \bar{\chi}^c_R \chi_R + \text{h.c.} \,, \\
\mathcal{L}_V = & -\dfrac{1}{4} A'^{\mu \nu} A'_{\mu\nu} - \dfrac{1}{2} \dfrac{\epsilon}{\cos{\theta_w}} B^{\mu \nu} A'_{\mu\nu} \, , \\
    \mathcal{L}_{S} = & (D^{\mu} S)^*(D_{\mu} S) + \mu_s^2 \abs{S}^2 - \lambda_s \abs{S}^4 -  \lambda_{hs} \abs{S}^2 \abs{H}^2 \, ,  
    \label{eq:Lscalar}
\end{align} 
 with the SM Higgs field denoted by $H$, the SM hypercharge gauge boson by $B$, the covariant derivitive by $ i D_{\mu} \phi = i \partial_{\mu} \phi - q_{\phi} g_{\chi} A'_{\mu} \phi$ and the charge conjugated field by $\psi^c = C \gamma_0^T \psi^*$ where $C$ is the charge conjugation matrix.  

The parameters in $\mathcal{L}_{\chi}$ can be considered to be real and positive by field redefinitions up to one complex phase which can be chosen to appear in the Dirac mass, $m_D = m_d e^{i\phi_d}$, with $m_d \geq 0$. In the following we will always assume that this phase is zero, corresponding to C conservation in the dark sector.\footnote{Allowing for a complex phase would introduce a $p$-wave contribution to the co-annihilation channel, which has little implications for the phenomenology studied here, since this interaction is relevant only at high energies, see~\cref{app:general-model}.}  The gauge sector parameters can also be taken to be real and positive by field and charges redefinitions. We require that the scalar field obtains a vev, which spontaneously breaks the $U'(1)$ symmetry. A complex phase of the dark vev $w$ can be removed by redefining the scalar field $S$. Without lost of generality, we take $q_{\chi} = 1$ since any other value can be absorbed in the definition of the gauge coupling constant $g_{\chi} \to g_{\chi}/q_{\chi} $.

After the $U(1)'$ symmetry breaking, the dark photon acquires a mass $m_{A'} = 2 g_\chi w$ and we obtain the fermion mass terms
\begin{equation}
    \mathcal{L}_\chi \supset - m_d \bar{\chi}_L \chi_R -  \frac{1}{2} m_L \bar{\chi}^c_L \chi_L -  \frac{1}{2}m_R  \bar{\chi}^c_R \chi_R + \text{h.c.} \, .
\end{equation} 
The Majorana masses are related to the Yukawa couplings and the scalar vev as $m_{L/R} = 2 y_{L/R} w$. 
After a rotation  of the fields $(\chi_L, \chi_R^c)$ with a mixing angle $\theta$ and
a phase-changing unitary transformation $\text{diag}(1,i)$ one finds two physical Majorana states, which we will denote by $\chi$ and $\chi^{\ast}$, with masses
\begin{align}
    m_{\chi^{(\ast)}}=  \sqrt{m_d^2+  w^2(y_R-y_L)^2} \mp w  (y_L+y_R) \, .
    \label{eq:mass_eigenstates}
\end{align} 
The state $\chi$ is the DM particle, whereas $\chi^\ast$ plays the role of the ``excited'' DM state, with $m_{\chi^\ast} > m_\chi$.\footnote{To obtain eq.~\eqref{eq:mass_eigenstates}, we have assumed $m_d \geq 2 w\sqrt{y_R y_L}$. If this inequality is violated, the phase-changing unitary transformation diag$(1,i)$ is unnecessary, which causes the co-annihilation channel to change from $s$-wave to $p$-wave (see~\cref{app:general-model} for details). }

The same rotation applied to the interaction part of the Lagrangian leads to diagonal and non-diagonal interactions between $\chi$ and $\chi^{\ast}$ with the $U'(1)$ gauge boson $A'$. These interactions can be characterised by the diagonal fine-structure constant $\alpha'_\text{el}$ and the off-diagonal one $\alpha'_\text{inel}$ given by\footnote{We use terms elastic and inelastic for the couplings definitions in order to make connection with the fact that niDM can be seen as a generalization of iDM models. }
\begin{align}\label{eq:alpha}
    \alpha'_\text{el} = \alpha' \cos^2{2\theta} \,, \qquad \alpha'_\text{inel} = \alpha' \sin^2{2\theta} \; .
\end{align}     
with $\alpha' = g_{\chi}^2 / (4\pi)$.\footnote{Note that in our model the diagonal couplings $\chi\chi A'$ and $\chi^\ast\chi^\ast A'$ are equal in norm and given by $|g_\chi\cos 2\theta|$. This is a consequence of our assumption of equal $U(1)'$ charges of $\chi_L$ and $\chi_R$ as required for anomaly cancellation.} The angle of the $\chi_{L/R}$ fields rotation $\theta$ is given by 
\begin{equation}\label{eq:rotationAngleTheta}
    \cos{2\theta} = - \dfrac{w (y_R - y_L)}{\sqrt{m_d^2 + w^2 (y_R - y_L)^2}} = - \dfrac{\delta_y \Delta_m}{(2+\delta_y)(2+\Delta_m)} \, ,
\end{equation} 
where in the last equality we have rewritten the equation by introducing the normalized dark left-right Yukawa (or Majorana mass) asymmetry 
\begin{align}\label{eq:delta_y}
\delta_y \equiv \frac{y_R - y_L}{y_L} = \frac{m_R - m_L}{m_L}    
\end{align}
and the relative mass splitting between the fermionic mass eigenstates 
\begin{align}\label{eq:Delta_m}
\Delta_m \equiv \frac{m_{\chi^{\ast}}-m_{\chi}}{m_{\chi}} \,.   
\end{align}
Without loss of generality we take $\delta_y>0$, which can always be achieved by an appropriate re-labeling of $\chi_L \leftrightarrow \chi_R^c$.

Since the symmetries of the model allow for a kinetic mixing $\epsilon$ between the dark $U'(1)$ group and the SM $U_Y(1)$ hyper-charge, the dark photon can interact with SM fermions via the photon or $Z$ boson portal. In the case of $m_{A'} \ll m_Z$, only the photon portal is relevant such that SM fermions $f$ obtain an effective $U'(1)$ charge given by the electromagnetic charge of the particle $Q_f$ times the negative of the kinetic mixing $\epsilon$.

To simplify our discussion, we assume that the scalar sector is irrelevant for the phenomenology. Given the tight experimental bounds on the Higgs portal coupling (see ref.~\cite{Ferber:2023iso} for a review) and the strong Yukawa suppression of the Higgs couplings to light SM particles ($e,\mu,\pi,p,$...), this is guaranteed to be the case for DM scattering. For the annihilation process $\chi \chi \to \phi \phi$ to be irrelevant for the relic density calculation, we require $m_s \gtrsim 1.5 \, m_\chi$, which can be achieved if the dark Higgs self-coupling is close to the perturbativity bound~\cite{Duerr:2016tmh}. However, since we will study large values of the dark gauge coupling, it is not possible with perturbative couplings to achieve $m_{s} \gtrsim m_{A'}$, and production of dark Higgs bosons might be relevant for accelerator searches such as the ones discussed in ref.~\cite{Duerr:2020muu}. We leave a detailed exploration of such signatures to future work. In other words, we only make use of the scalar sector to consistently implement the $U(1)'$ symmetry breaking. The scalar Lagrangian $\mathcal{L}_S$ given in \cref{eq:Lscalar} serves just as a concrete example and the subsequent discussion does not depend on its specific form.

In summary, the independent parameters relevant for the DM phenomenology can be chosen as
\begin{align}
    \epsilon, \, \alpha' ,\, \delta_y, \, \Delta_m, \, m_\chi, \, m_{A'} \,.
\end{align}
We highlight in particular the left-right asymmetry parameter $\delta_y$ which is absent in iDM models. Indeed, setting $\delta_y = 0$ leads to $\theta = \pi/2$ and hence $\alpha_\text{el} = 0$ and $\alpha'_\text{inel} = \alpha'$, corresponding to purely off-diagonal interactions. The main focus of our work will be to study how the phenomenology of the model changes for $\delta_y > 0$. We note that if $y_L \ll y_R$ it is possible for $\delta_y$ to be much larger than unity without violating unitarity or perturbativity.

To simplify the parameter space further, we follow a common approach in the literature and fix some of the model parameters to specific benchmark values:
\begin{align}
    \alpha'= 0.5 \,,\qquad m_{A'} = 3 m_\chi \,.
\end{align}
We note that these values are optimistic in the sense that they enhance experimental sensitivity and relax the relic density constraint.
As we will see below, for these values of $\alpha'$ and $m_{A'} / m_\chi$ the most interesting ranges for the DM mass are  200~MeV~$\lesssim m_\chi \lesssim 500$~MeV and 3~GeV~$\lesssim m_\chi \lesssim 8$~GeV.

\begin{figure}
     \centering
     \includegraphics[width=0.85\textwidth]{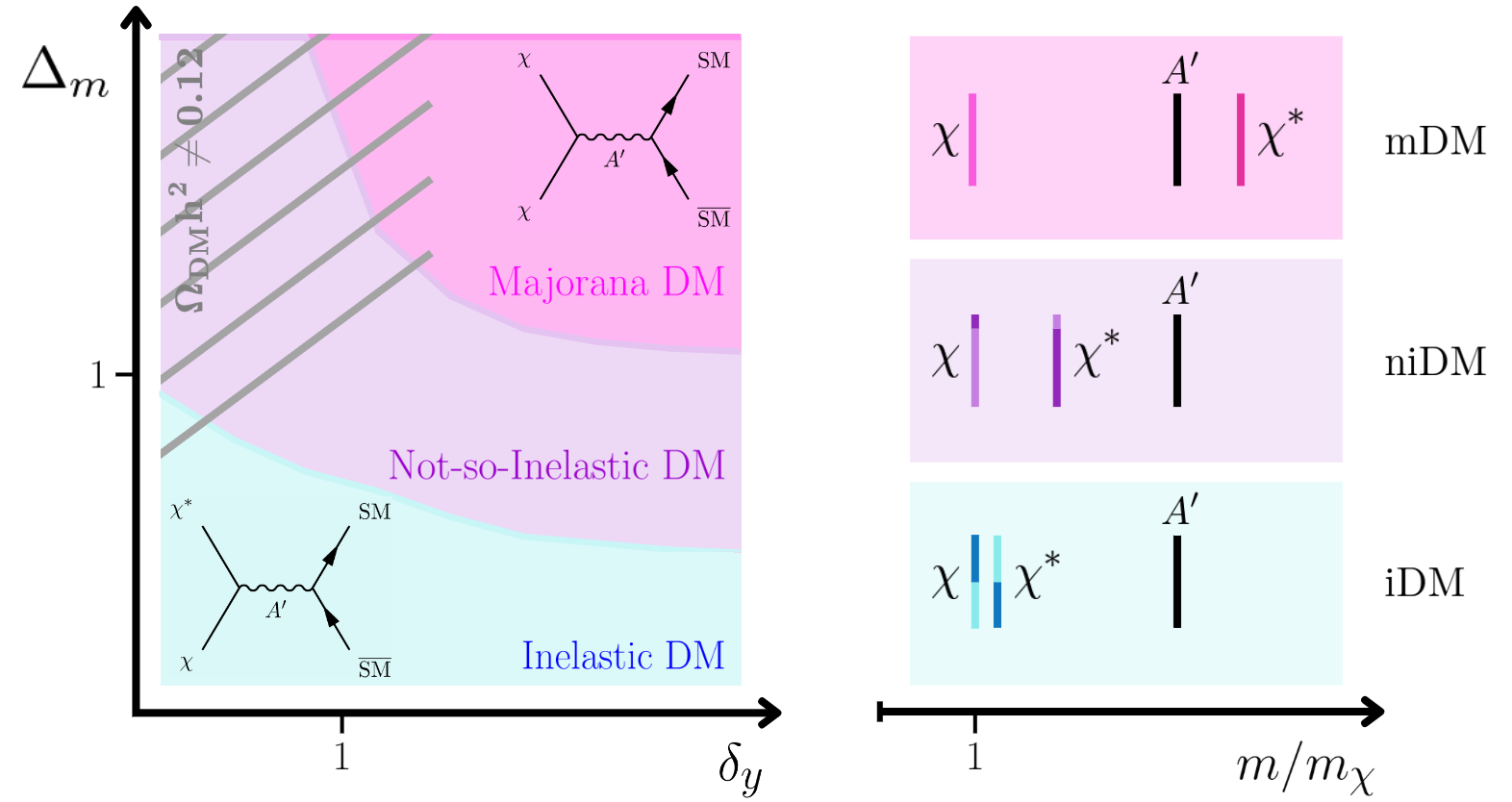}
        \caption{Schematic illustration of the particle spectrum and the continuous transition from inelastic DM (iDM) over not-so-inelastic DM (niDM) to Majorana DM (mDM). The left panel shows the effect of the mass splitting $\Delta_m$ and the Yukawa asymmetry $\delta_y$, see \cref{eq:Delta_m,eq:delta_y}, where we indicate also the relevance of elastic (diagonal) and inelastic (off-diagonal)  interactions. The right panel shows the typical relative mass spectrum of $\chi,\chi^\ast$ and the dark gauge boson $A'$ for different regimes of the mass splitting $\Delta_m$.}
        \label{fig:spectrum}
\end{figure}

To conclude this section, let us consider in more detail the role played by the two parameters $\delta_y$ and $\Delta_m$ defined in \cref{eq:delta_y,eq:Delta_m}, respectively. The various regimes are illustrated in \cref{fig:spectrum}:
\begin{itemize}
\item \textit{iDM:} For $\Delta_m\ll 1$ the two Majorana fermions combine to a  pseudo-Dirac particle with a mass $m_d$ and small mass splitting. Moreover, the diagonal couplings of $\chi$ and $\chi^\ast$ to $A'$ are suppressed compared to the off-diagonal interaction irrespective of the value of $\delta_y$. This corresponds to the usual inelastic DM model. It arises naturally if there is a separation of scales, $m_d \gg m_{L/R}$, related to $U(1)'$ conserving and breaking new physics. 
\item \textit{mDM:} For $\Delta_m \gg 1$ (i.e., $m_{\chi^{\ast}}\gg m_{\chi}$) and $\delta_y \gtrsim 2$ we have the hierarchies $m_L\ll m_R$ and $m_d \ll  m_R$, which leads to two separated Majorana particles with a seesaw type spectrum: $m_\chi \simeq m_d^2/m_R - m_L \ll m_{\chi^\ast} \simeq  m_R$. At the same time, the mixing angle $\theta$ can be very different from $\pi/2$, such that there are sizeable diagonal interactions. As a result, the heavier state becomes essentially irrelevant, and we are left with a simple realization of a Majorana DM model. 
\item \textit{niDM:}
A third regime exists which combines features from both cases, in particular a DM relic abundance calculation similar to mDM but experimental signatures more closely related to iDM. This regime corresponds to the region of $\Delta_m \sim 1$ and $\delta_y \gtrsim 1$ and features a mix of elastic (diagonal) and inelastic (off-diagonal) DM interactions.  This is the regime we are most interested in this study and which we call not-so-inelastic Dark Matter (niDM).
\end{itemize}
We note that there exists a fourth regime, corresponding to $\Delta_m \gtrsim 1$ and $\delta \lesssim 1$, where the mass splitting is large but diagonal interactions are suppressed. In this regime, it is typically impossible to reconcile the relic density requirement with experimental constraints, so that we will not consider it in detail in the present work.
}

To conclude this section, we point out that Big Bang Nucleosynthesis places a robust lower bound of approximately $m_\chi \gtrsim 10 \, \mathrm{MeV}$ on the mass of any DM particle in thermal equilibrium~\cite{Sabti:2019mhn}. In the niDM regime, where $\Delta_m \sim 1$, the decay $\chi^\ast \to \chi e^+ e^-$ is therefore always kinematically allowed. As we will show in the following section, this implies that the excited state is always short-lived on cosmological timescales.

{
\section{Excited Dark Matter decays}
\label{sec:decays}

The fact that the dark sector in the niDM model contains excited DM particles that may decay into SM states is a key factor for its phenomenology. First of all, we need to ensure that the excited state is short-lived compared to cosmological timescales, such that bounds on decaying DM components are evaded. For the collider phenomenology, we are furthermore interested in the branching ratios for the various final states, which determine the resulting signatures.

If the interactions between the two sectors are dominantly mediated by the dark photon, one finds that above the decay threshold, i.e.\ for $\Delta_m m_{\chi} > 2m_f$,  the spin-averaged square of the decay amplitude for $\chi^{\ast} \to \chi f\Bar{f}$ is approximately given by \begin{equation} \label{eq:ChiAstDecaysIntoFermions}
    \overline{ \lvert \mathcal{M} \lvert^2 }(\chi^{\ast} \to \chi \, f\bar{f})  \approx  \alpha_{\text{dec}} Q_f^2 \left[4 \Delta _m \left(x_{\chi }-1\right)-2 x_f x_{\chi }-x_f^2-2 \left(x_{\chi }-1\right)^2\right]
\end{equation} with
\begin{equation}
    \alpha_{\text{dec}}=  64 \pi ^2 \epsilon ^2  \alpha  \,  \alpha'_{\text{inel}} \frac{m_{\chi }^4}{m_{A'}^4} \, .
\end{equation} We adopt the notation $x_{\chi} =m_{\chi P}^2/m_{\chi}^2$ and $x_P = m_{P\bar{P}}^2/m_{\chi}^2$ for final state particles $\chi P \Bar{P}$ where $m_{P_1 P_2}$ is the invariant mass of the pair of particles $P_1 P_2$. From the amplitude above, we find the partial decay width \begin{equation}
    \Gamma(\chi^{\ast} \to \chi \, f\bar{f}) \approx N_c Q_f^2   \dfrac{\alpha_{\text{dec}}}{240\pi^3}  \Delta_m ^5 m_{\chi } \, ,
\end{equation} where $N_c = 3$ for quarks and $N_c = 1$ for leptons. In all equations of this section, we assume for compactness that the SM fermions are light compared to the mass splitting and the mass splitting is small compared to both the dark photon mass and the DM mass: $m_f \ll \Delta_m m_{\chi} \ll m_\chi$ and $m_{f\bar{f}}^2 \leq \Delta_m^2 \ll m_{A'}^2$. However, we keep the full mass dependence for the numerical results presented below.

Considering only the above perturbative decays into fermions, we can estimate the lifetime of the excited state as \begin{equation}
    \tau_{\chi^{*}} \sim  10^{-15}\text{ s}\left(\dfrac{0.5}{\Delta_m}\right)^5 \dfrac{0.5}{\alpha'_{\text{inel}}}  \left(\dfrac{0.01}{\epsilon}\right)^2 \dfrac{500 \text{ MeV}}{m_\chi} \left(\frac{m_{A'}}{m_{\chi }}\right)^4 \, .
\end{equation}  
The above estimate shows that we expect extremely short-lived excited DM states in the niDM regime. Therefore, we can completely neglect the $\chi^\ast$ abundance after thermal decoupling and only focus on searches for its decays in high energy experiments.

In order to correctly describe non-perturbative QCD effects related to $\chi^\ast$ decays into hadrons arising below $\Delta_m m_{\chi}\approx 2$~GeV, we adopt the data-driven approach given in appendix~A of ref.~\cite{Duerr:2020muu}. It makes use of the measured cross section ratio $R(s) \equiv \sigma(e^- e^+ \to \textit{hadrons})/\sigma(e^- e^+ \to \mu^- \mu^+)$~\cite{PhysRevD.98.030001} at collisions with center-of-mass energy $\sqrt{s}$ in order to rescale the (analytically calculated) $ \Gamma(\chi^{\ast} \to \chi \, \mu^- \mu^+)$ into $ \Gamma(\chi^{\ast} \to \chi +\textit{hadrons})$. This prescription is used to calculate the total decay width into hadrons, but it does not provide partial decay widths for exclusive final states or their kinematics. Since we will be interested in these aspects for the simulation of $\chi^{\ast}$ decays, we also consider vector meson dominance (VMD)~\cite{Benayoun:1998ss,Bondarenko:2018ptm} theory to compute the relevant matrix elements for exclusive decay modes. We focus only on 3-body channels, namely \begin{align}\label{eq:ChiAstDecaysIntoPions}
    &\overline{ \lvert \mathcal{M} \lvert^2 }(\chi^{\ast} \to \chi \, \pi^+ \pi^-) = \frac{\alpha_{\text{dec}}}{4} G(x_{\chi }, x_\pi) \big\lvert F_{\pi}(m_{\pi\pi}^2)  \big\lvert^2 \\ \label{eq:ChiAstDecaysIntoKaons}
    &\overline{ \lvert \mathcal{M} \lvert^2 }(\chi^{\ast} \to \chi \, KK) =\frac{\alpha_{\text{dec}}}{36} m_{\rho}^4\frac{x_K (4 x_{\chi }-\left(\Delta _m+2\right)^2)+\left(\left(\Delta _m+2\right) \Delta _m-2 x_{\chi }+2\right)^2}{(m_{\phi }^2-x_K m_{\chi }^2)^2+\Gamma _{\phi }^2 m_{\phi }^2} 
\end{align} where
\begin{equation}
     G(x_{\chi }, x_\pi) = 4 \left[(x_{\chi }-1) (x_\pi+x_{\chi }-1) - \Delta _m (x_\pi+2 x_{\chi }-2)\right] -\Delta _m^2 (x_\pi+4 x_{\chi }-8) 
\end{equation}
and $F_{\pi}(q^2)$ is the pion electromagnetic form factor obtained by VMD fitting the experimental data~\cite{BaBar:2012bdw} (see also ref.~\cite{Bondarenko:2018ptm}). Here, $KK$ stands for both $K^+K^-$ and $K^0\Bar{K}^0$ neglecting their mass differences. The  analytical expressions in~\cref{eq:ChiAstDecaysIntoPions,eq:ChiAstDecaysIntoKaons} are only valid above the respective decay thresholds and for $m_{A'}^2, m_\chi^2 \gg \Delta_m^2 m_\chi^2$. 

In in~\cref{fig:ChiAstBranchingRatios} we show the branching fractions for the various leptonic and hadronic final states. 
Unlike for the decays of dark photons~\cite{Ilten:2018crw}, the hadronic decay modes of $\chi^{\ast}$ do not exhibit any sharp resonances since one integrates over the invariant mass of the hadronic decay products. 
By comparing the branching fractions into pions and kaons with the total branching fraction into hadrons, we can conclude that additional hadronic channels remain sub-dominant (with contributions below 35\%) up to $\sim$2~GeV. For higher masses one could in principle calculate the decay width into free quarks following~\cref{eq:ChiAstDecaysIntoFermions} and perform the hadronisation numerically. However, missing energy searches at current and past electron-positron colliders (see \cref{sec:5p3missingEnergySearchesAtColliders}) only cover the range $m_{A'} \lesssim 8$~GeV, such that these large mass splittings are irrelevant in the context of our benchmark scenario $m_{A'} = 3 m_{\chi}$. 

\begin{figure}
     \centering
     \includegraphics[width=0.55\textwidth]{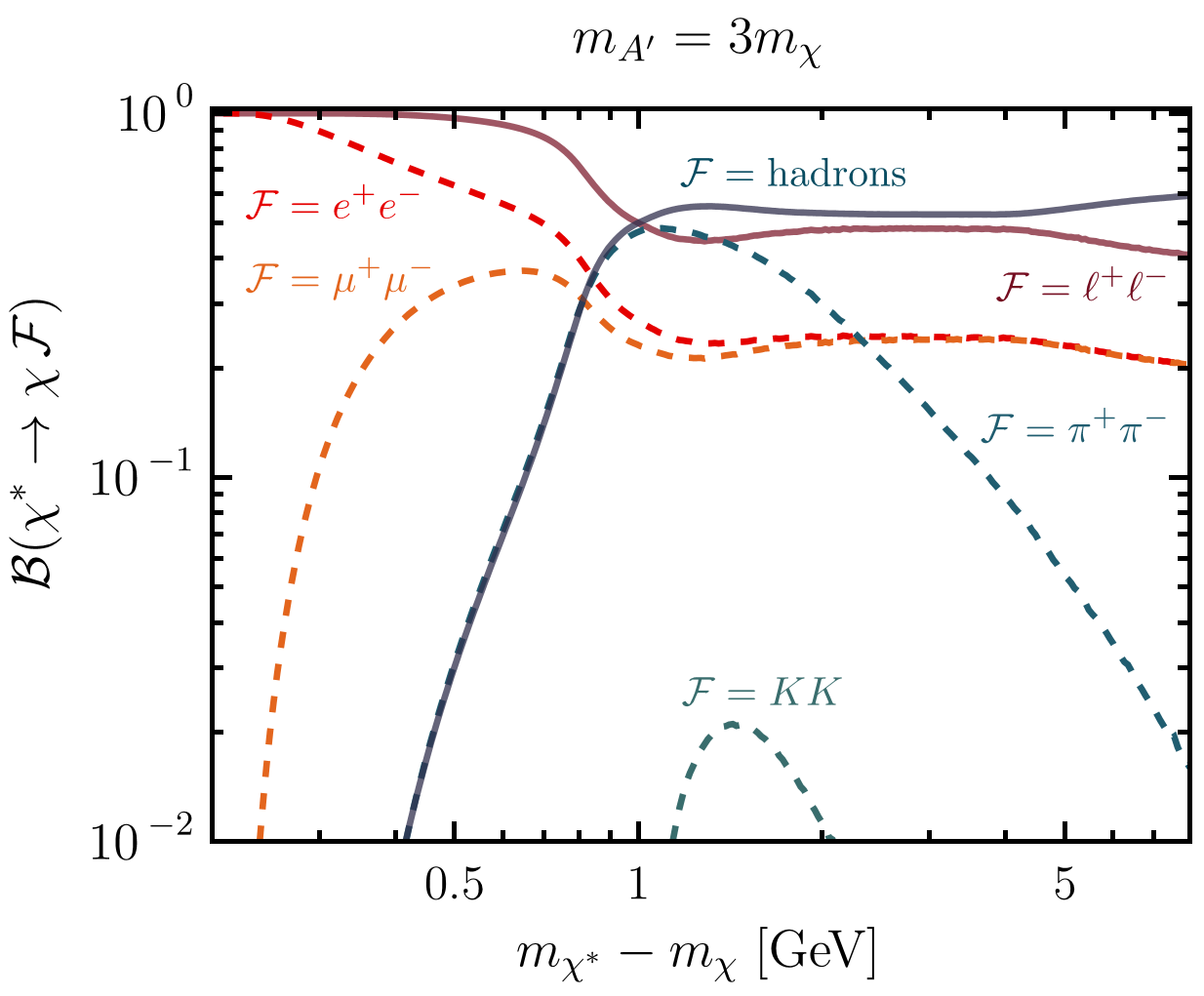}
        \caption{Decay branching fractions for the excited DM state $\chi^\ast$ in the niDM model as a function of the mass splitting between dark fermionic states.}
        \label{fig:ChiAstBranchingRatios}
\end{figure}

 We emphasize that while we have fixed $m_{A'} = 3 m_\chi$ in \cref{fig:ChiAstBranchingRatios}, the branching ratios are nearly independent of this choice for sufficiently heavy dark photons. This is because $m_{A'}$  only appears in the dark photon propagator, $1/(m_{f\bar{f}}^2 - m_{A'}^2)$, and because $m_{f\bar{f}}^2 \leq \Delta_m^2 m_\chi^2$. Hence, as long as $m_{A'}^2 \gg \Delta_m^2 m_\chi^2$, one can simply approximate the propagator by $-1/m_{A'}^2$ which cancels out when computing decay width fractions. Moreover, as long as the decay modes $\chi^\ast \to 3 \chi$ and $\chi^\ast \to \chi A'$ are kinematically closed, the $\chi^\ast$ branching fractions only depend on the absolute mass splitting, $ \Delta_m  m_\chi = m_{\chi^\ast} - m_\chi$, which   implies that the results from \cref{fig:ChiAstBranchingRatios}
 can be used for both iDM models with $\Delta_m \ll 1$ and niDM models with $\Delta_m \approx 1$.
}

{\section{Relic abundance}
\label{sec:relic}
In this section we present the calculation of the DM relic abundance in the niDM model, focusing especially on the interplay between the left-right asymmetry parameter $\delta_y$ and the dark-fermion mass splitting $\Delta_m$ defined in \cref{eq:delta_y,eq:Delta_m}, respectively.

We compute the DM relic abundance using the public tool \texttt{MicrOmegas v5.3.41}~\cite{Alguero:2022inz} inserting a \texttt{CalcHEP} model file~\cite{Pukhov:2004ca} written with the help of the \texttt{FeynRules v2.3.49 Mathematica} package~\cite{Alloul:2013bka}. To account for non-perturbative QCD effects, following usual methods~\cite{Izaguirre:2015yja,Berlin:2018bsc}, we have made some modifications to the annihilation cross sections calculated in MicrOmegas.  First of all, for $\chi^{(\ast)} \chi^{(\ast)}$ annihilations with center-of-mass energy, $\sqrt{s} \approx m_{\chi^{(\ast)}} + m_{\chi^{(\ast)}}$, below the pion mass, we have set the annihilation cross section into light quarks ($d$, $u$, $s$, $c$) to zero. For energies above the pion mass, we replace the annihilation cross section into light quarks by the (analytically calculated) annihilation cross section into a muon pair rescaled by the measured cross section ratio $R(s)$~\cite{PhysRevD.98.030001}, i.e.,
\begin{equation} \label{eq:RfactorHadronsCrossSect}
    \sigma(\chi^{(\ast)} \chi^{(\ast)}  \to \textit{hadrons}) = R(s) \, \sigma(\chi^{(\ast)} \chi^{(\ast)} \to \mu^- \mu^+) \, ,
\end{equation} see appendix A of ref.~\cite{Duerr:2020muu} for further details.
Note that we only consider $m_{A'} > m_\chi$ in the present work, such that pure dark sector annihilations $\chi\chi \to A'A'$ are kinematically forbidden. Such annihilations would lead to a secluded dark sector, which can have tiny couplings to the SM. Furthermore, dark sector co-annihilations $\chi\chi^\ast \to A'A'$ are always unimportant: our benchmark choice $m_{A'} = 3m_\chi$ would require $\Delta_m>4$ to allow for co-annihilations, which corresponds to a very strong Boltzmann suppression of the $\chi^\ast$ abundance, rendering co-annihilations irrelevant.

\begin{figure}
        \centering
     \begin{subfigure}{0.42\textwidth}
        \centering
        \includegraphics[width=.99\linewidth]{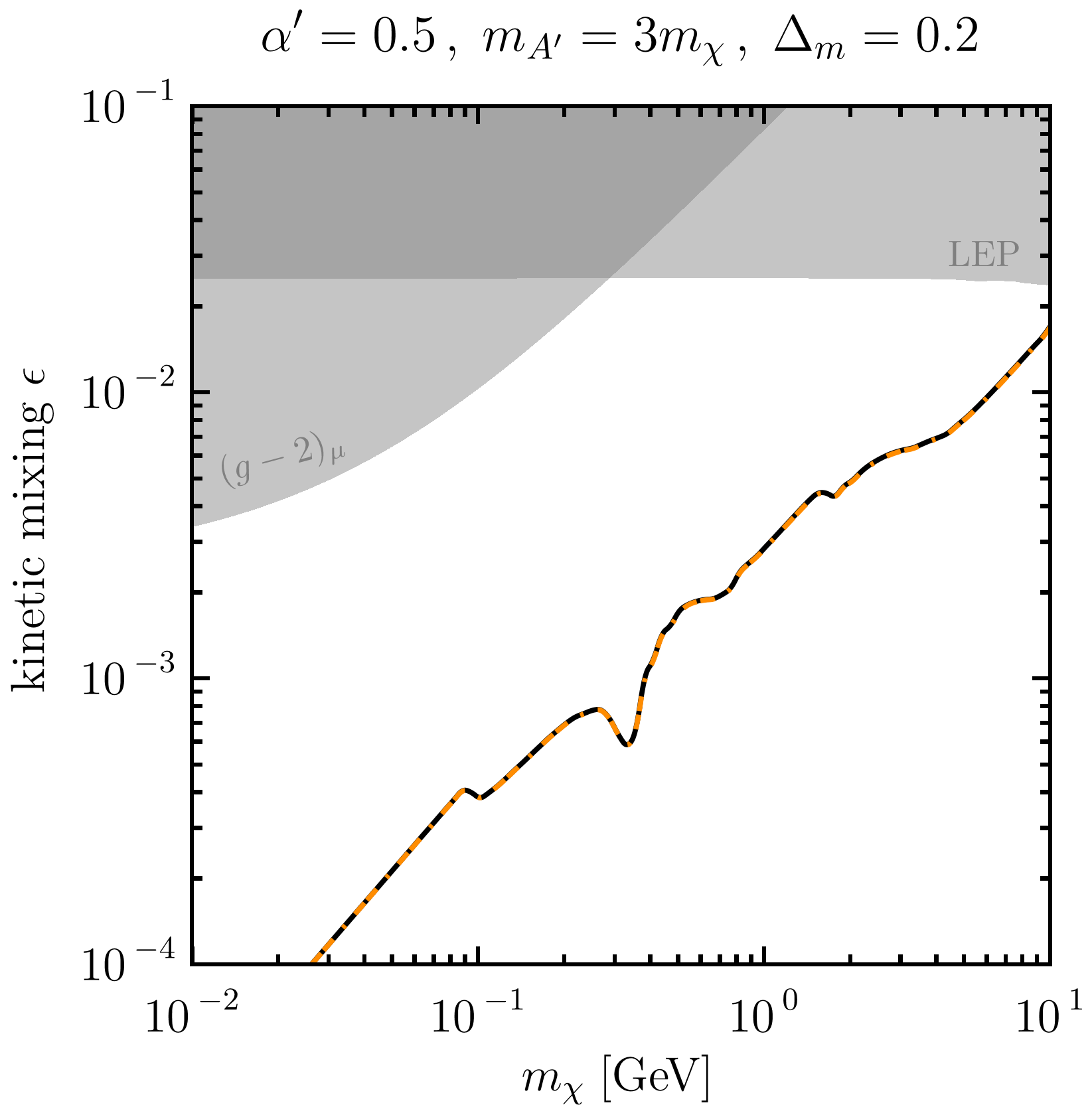}
     \end{subfigure}
     \hspace{0.05cm}
      \begin{subfigure}{0.42\textwidth}
        \centering
        \includegraphics[width=.99\linewidth]{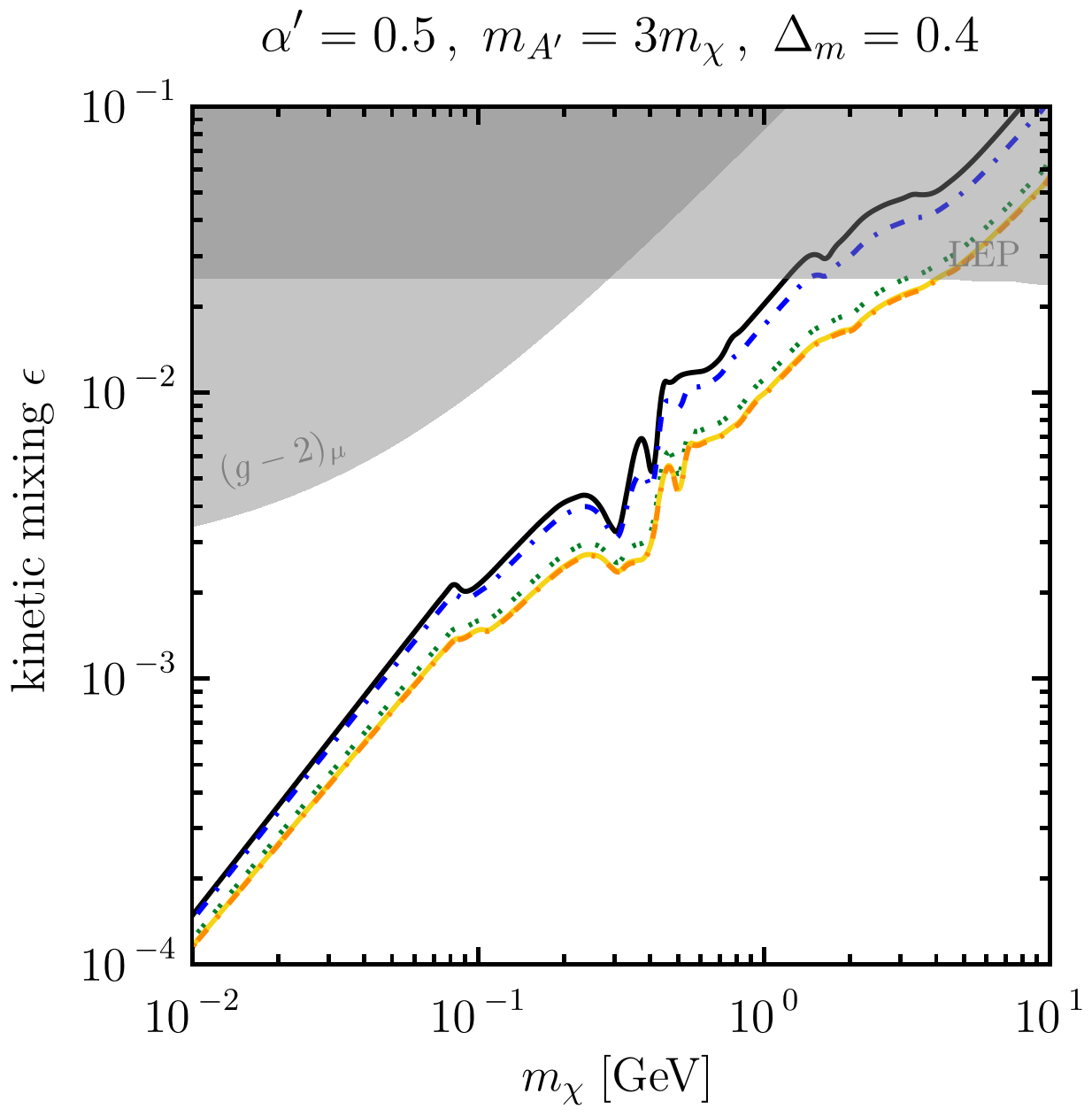}
     \end{subfigure}
      \begin{subfigure}{0.42\textwidth}
        \centering
        \includegraphics[width=.99\linewidth]{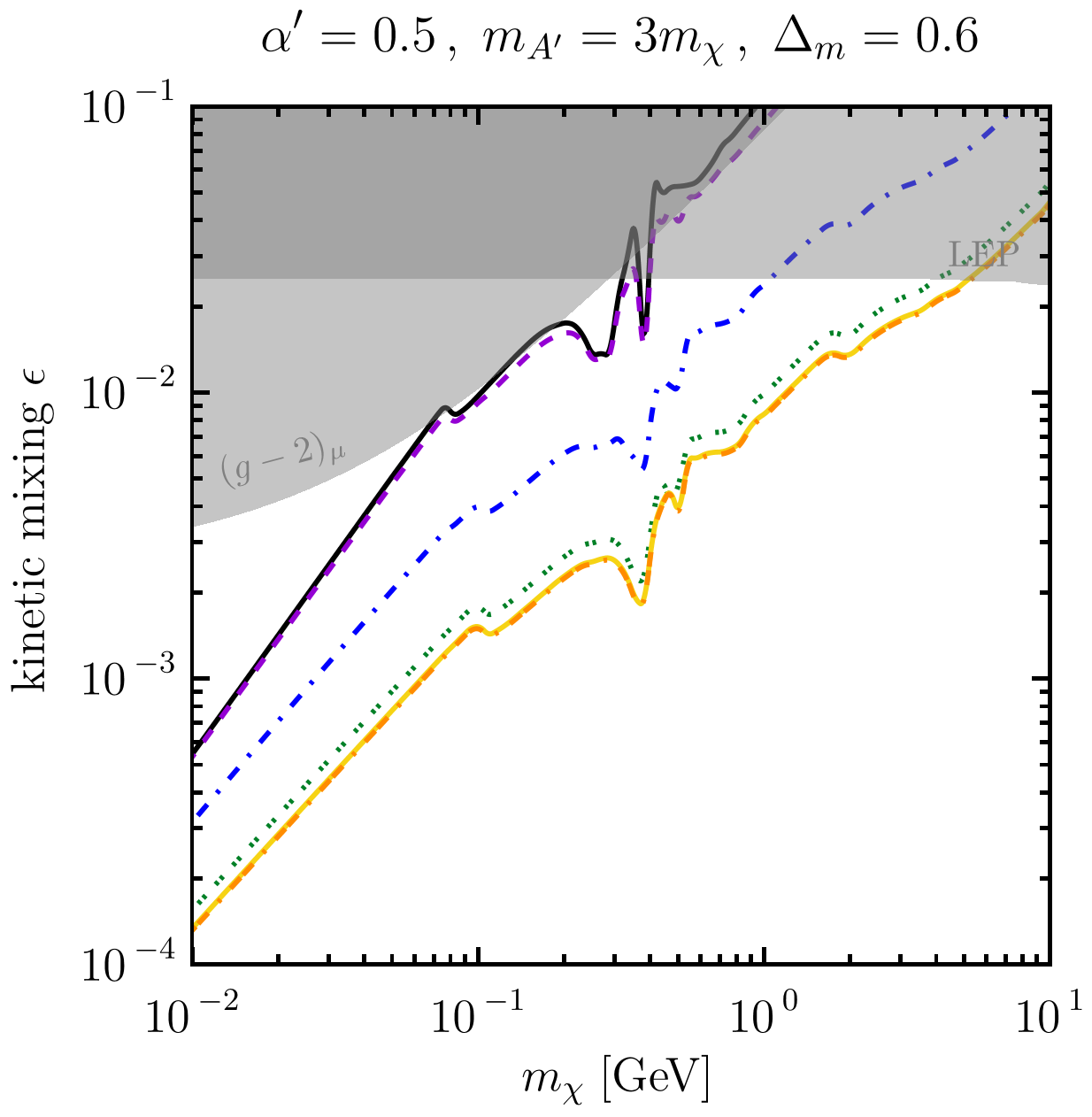}
     \end{subfigure}
     \hspace{0.05cm}
      \begin{subfigure}{0.42\textwidth}
        \centering
        \includegraphics[width=.99\linewidth]{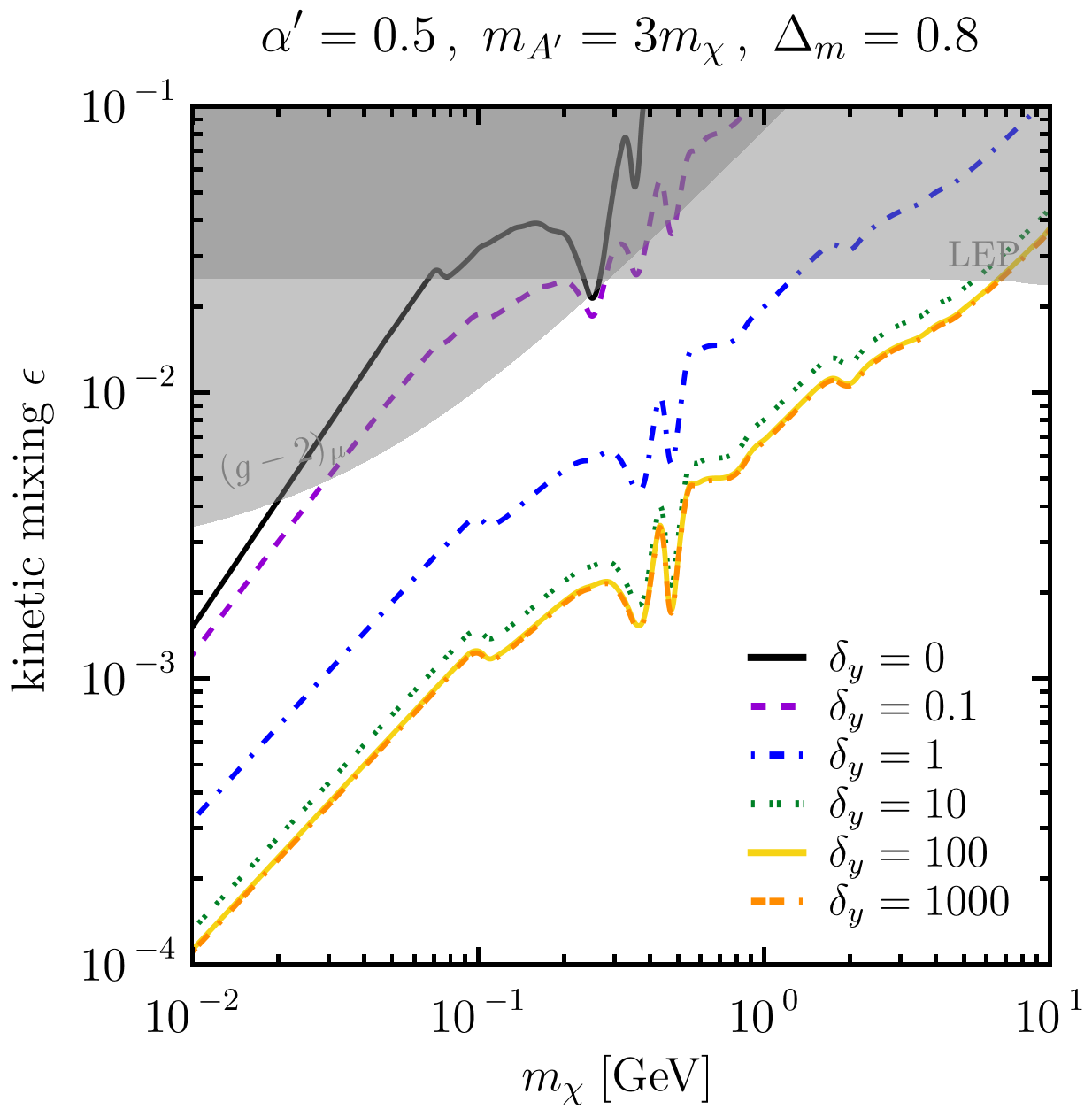}
     \end{subfigure}
        \caption{Values of the kinetic mixing angle $\epsilon$ which reproduce the observed DM relic abundance $\Omega_{\rm DM}h^2 = 0.12$ as a function of the DM mass. The four panels correspond to different normalized mass splittings $\Delta_m=(0.2,0.4,0.6,0.8)$. The various curves in each panel correspond to different values of the dark left-right asymmetry~$\delta_y$, with the iDM limit~($\delta_y = 0$) shown in black. Gray shaded regions correspond to model-independent bounds on the kinetic mixing $\epsilon$ (see text for details).}
        \label{fig:relicAbundance2D}
\end{figure}

\Cref{fig:relicAbundance2D} shows the value of the kinetic mixing parameter $\epsilon$ required to reproduce the observed DM abundance as a function of the DM mass for different values of the left-right asymmetry parameter $\delta_y$ ranging from zero (i.e., the classical iDM limit) up to 1000. The plots also include model-independent bounds on the kinetic mixing~$\epsilon$ from the measurement of the muon anomalous magnetic dipole moment~\cite{Pospelov:2008zw} and from electroweak precision measurements at LEP~\cite{Hook:2010tw}. These bounds arise because a dark photon with a kinetic mixing can appear virtually in SM process and modify the corresponding observables.

As visible from the figure, the effect of $\delta_y$ strongly depends on the value of the mass splitting $\Delta_m$: for small values, $\Delta_m \lesssim 0.2$ (upper left panel), DM freeze-out is essentially independent of $\delta_y$, because the DM annihilation is always dominated by the off-diagonal coupling as in the conventional iDM scenario. This finding is a result of two effects: First of all, if $\chi^\ast$ is not much heavier than $\chi$, the Boltzmann suppression of both states is comparable and hence both states can participate in annihilation process. Furthermore, for small values of $\Delta_m$ the mixing angle $\theta$, and hence the ratio $\alpha_{\rm el}/\alpha_{\rm inel}$, does not depend strongly on $\delta_y$, see \cref{eq:rotationAngleTheta}.

However, for mass splittings $\Delta_m\gtrsim 0.6$ (bottom row), the Boltzmann suppression of the heavier state becomes relevant and large values of $\delta_y$ significantly enhance the DM annihilation rate, because the relative importance of diagonal interactions increases, c.f.~\cref{eq:alpha,eq:rotationAngleTheta}. As a result, the required values of $\epsilon$  can be lowered by more than one order of magnitude compared to the case with $\delta_y = 0$. We note also that for $\delta_y \gg 1$ the effect saturates, as $\cos 2\theta \propto \delta_y/(2+\delta_y) \to 1 - 2/\delta_y $.  

\begin{figure}
     \centering
     \begin{subfigure}{0.408611615245\textwidth}
        \centering
        \includegraphics[width=.99\linewidth]{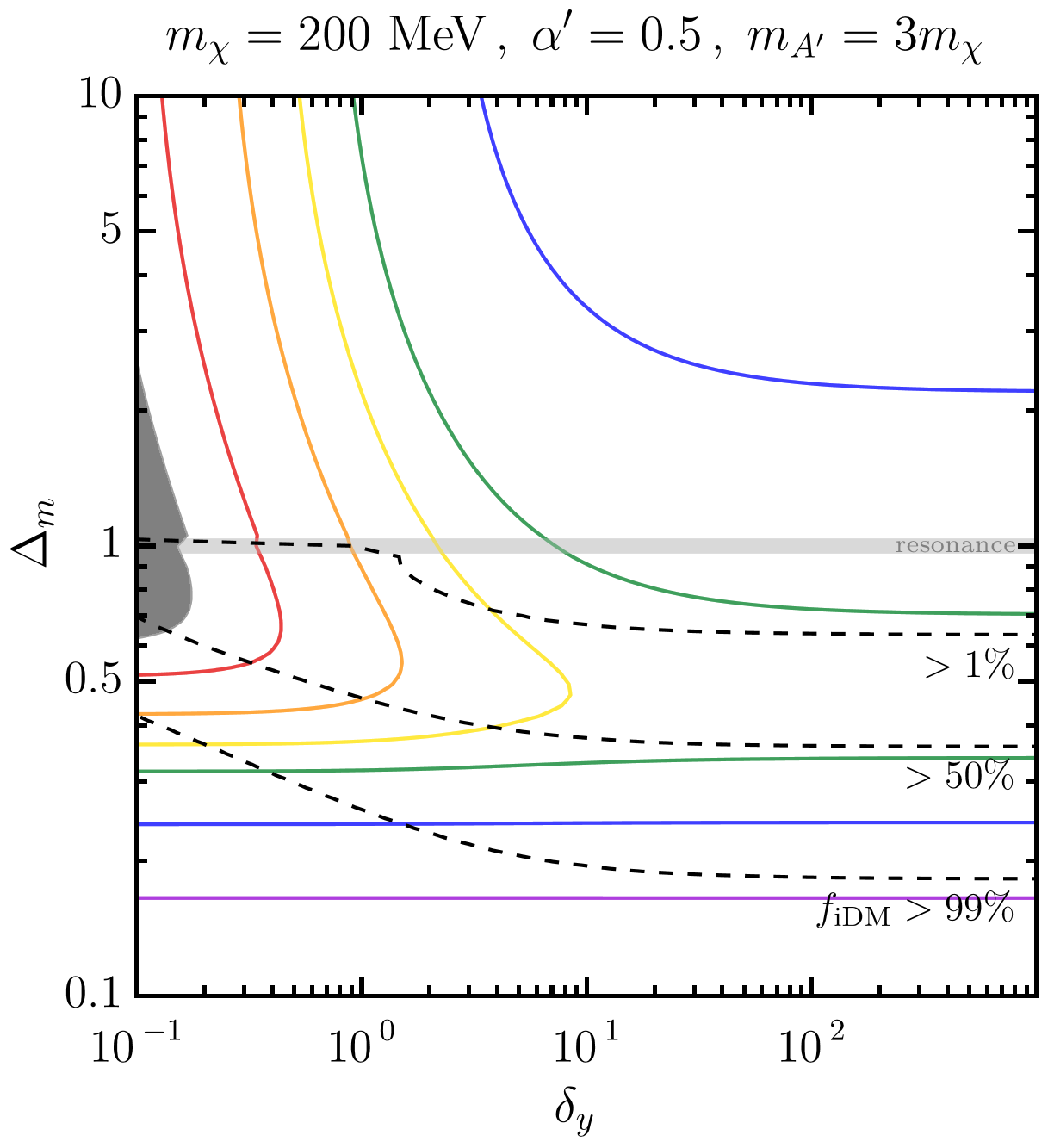}
     \end{subfigure}
     \hspace{0.05cm}
      \begin{subfigure}{0.516388384755\textwidth}
        \centering
        \includegraphics[width=.99\linewidth]{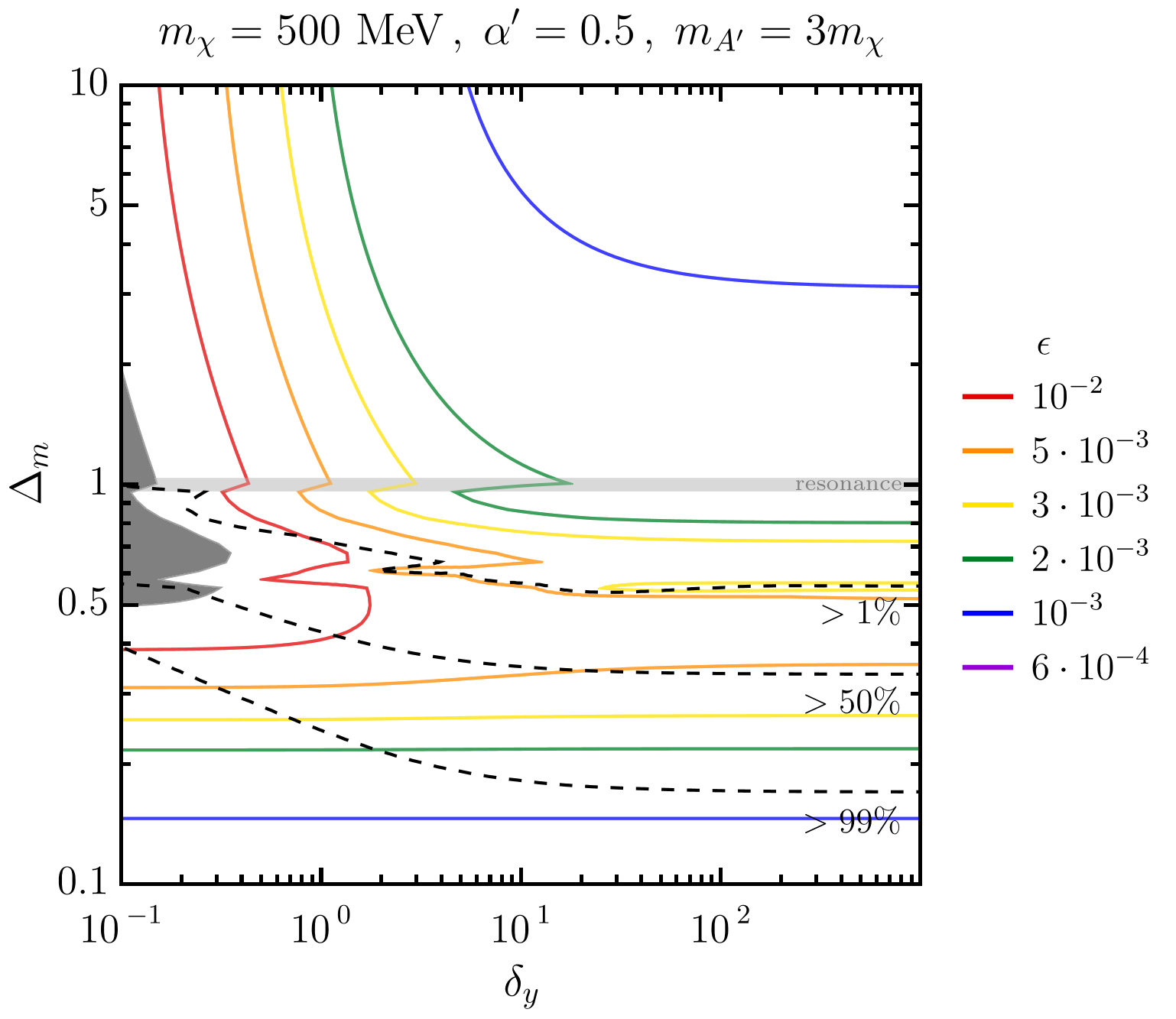}
     \end{subfigure}
        \caption{Contours of the kinetic mixing parameter $\epsilon$ reproducing the observed relic abundance $\Omega_{\rm DM}h^2 = 0.12$ as a function of the dark left-right asymmetry~$\delta_y$ and the normalized mass splitting~$\Delta_m$. The two panels correspond to different DM masses $m_{\chi}=(200,500)$~MeV. Dashed black lines show the relative contribution of the co-annihilation channel to the DM relic abundance (see text for details). The horizontal light gray band indicates when the co-annihilation channel becomes resonant due to $m_{\chi}+m_{\chi}^{\ast} = m_{A'} \pm 1\%$. Dark gray shaded regions are excluded by model-independent bounds on $\epsilon$.}
        \label{fig:relicAbundance3D}
\end{figure}

The features discussed above are further elucidated in \cref{fig:relicAbundance3D}, where we introduce plots in the $(\delta_y,\Delta_m)$ plane for two reference DM masses, $m_\chi=200\,\mathrm{MeV}$ (left) and $500\,\mathrm{MeV}$ (right). For each point in the $(\delta_y,\Delta_m)$ space we calculate the value of $\epsilon$ required to obtain the correct relic DM abundance, which are shown as coloured curves in the figure. We confirm that in the pseudo-Dirac limit ($\Delta_m \lesssim  0.2$) the DM abundance becomes independent of $\delta_y$, whereas for $\Delta_m \gtrsim 0.4$ the required value of $\epsilon$ becomes smaller for increasing $\delta_y$.

The black-dashed curves in \cref{fig:relicAbundance3D} indicate the relative contribution of off-diagonal interactions to the relic DM abundance, $f_{\rm iDM}$, defined following ref.~\cite{Belanger:2006qa}: In the so-called freeze-out approximation, the inverse total DM abundance $1/(\Omega_{\rm DM}h^2)$ is proportional to an integral of the velocity averaged total annihilation cross section, which can be split into elastic (diagonal) and inelastic (off-diagonal) contributions, 
\begin{align}
\frac{1}{\Omega_{\rm DM}} = 
\frac{1}{\Omega_{\rm DM}^\text{el}}
+\frac{1}{\Omega_{\rm DM}^\text{inel}} \,.
\end{align}
Then we define the relative contribution of the off-diagonal interactions as 
\begin{align}
    f_{\rm iDM} \equiv \frac{\Omega_{\rm DM}}{\Omega_{\rm DM}^\text{inel}} \,.
\end{align}
As visible in the figure, the niDM regime corresponds to iDM fractions of $f_{\rm iDM}\lesssim 50\%$, confirming that the diagonal coupling plays an important role in this regime. We note that curves of constant $f_{\rm iDM}$ do not follow exactly curves of constant $\alpha'_\text{inel}/\alpha'=\sin^22\theta$. The reason is that for sizeable mass splittings $\Delta_m$ the Boltzmann suppression of $\chi^\ast$ relative to $\chi$ affects the annihilation rate for the co-annihilation channel. 
}

{
\section{(In)Direct detection}

\label{sec:detection}

For the mass splittings that we consider here, inelastic DM scattering in direct detection experiments is kinematically strongly suppressed and therefore irrelevant.\footnote{The loop-induced elastic scattering process considered in ref.~\cite{Bell:2018zra} is kinematically allowed, but the cross section is too small to be obervable even in future direct detection experiments.} However, in the niDM regime sizeable elastic (diagonal) couplings are also present, which give rise to the effective low-energy interactions
\begin{equation}
\mathcal{L}_\text{eff} = \sum_{f} \frac{Q_f}{\Lambda^2} (\bar\chi \gamma_\mu \gamma_5\chi)(\bar f\gamma^\mu f) \; ,
\label{eq:Leff}
\end{equation}
where $\Lambda = m_{A'} / \sqrt{g_\chi \cos 2\theta \, \epsilon e }$.
The $\gamma^5$ in the first term is a result of the Majorana nature of the ground state and leads to a cross section that vanishes in the non-relativistic limit. Nevertheless, the absence of a $\gamma^5$ in the second term implies that DM-nucleus  scattering receives a coherent enhancement, which partially compensates for the velocity suppression. 

To calculate the sensitivity of direct detection experiments for our model, we first of all need to map the effective interaction from eq.~\eqref{eq:Leff} to the non-relativistic effective theory of DM-nucleon interactions introduced in ref.~\cite{Fitzpatrick:2012ix}. Using the public tool DirectDM~\cite{Bishara:2017nnn} with our model defined at the scale $\mu_c = 2$~GeV, we find that the niDM model introduces the effective operators $\mathcal{O}_8^N \propto \mathbf{v}_{\perp}$ and $\mathcal{O}_9^N  \propto \mathbf{q}$ with coefficients
\begin{equation}
c_8^p = \frac{2}{\Lambda^2} \, , \quad c_8^n = 0 \, , \quad c_9^p = \frac{5.586}{\Lambda^2} \, , \quad c_9^n = -\frac{3.826}{\Lambda^2} \; .
\end{equation}
following the conventions of ref.~\cite{Bishara:2017nnn}. 

As described in ref.~\cite{GAMBIT:2021rlp}, these coefficients can be directly passed to the public code DDCalc~\cite{GAMBIT:2018eea}, where likelihood functions for various direct detection experiments have been implemented. For the local DM distribution, we use results from ref.~\cite{Reynoso-Cordova:2024xqz} which take into account the influence of the Large Magellanic Cloud (LMC) on the DM velocity distribution (see \cref{app:DDupdates} for further details). We find that, depending on the choice of $\delta_y$ and $\Delta_m$, existing direct detection experiment, in particular LZ~\cite{LZCollaboration:2024lux}, already probe allowed regions of parameter space at the upper end of the mass range that we consider ($m_\chi \sim 10 \, \text{GeV})$ . We have also implemented the projected sensitivities for SuperCDMS~\cite{SuperCDMS:2016wui} and DARWIN~\cite{Schumann:2015cpa} in DDCalc, which are found to probe values of $\epsilon \cos 2 \theta$ as small as $10^{-3}$ across a wider mass range. These experiments will thus provide  complementary information on models of niDM.

For indirect detection, off-diagonal interactions are once more irrelevant, since the excited states $\chi^{\ast}$ are too heavy to be produced in DM scattering and too short-lived to be efficiently produced in astrophysical systems~\cite{Baryakhtar:2020rwy,Emken:2021vmf,Heeba:2023bik}. There is also no cosmological abundance of excited states, since the exponential suppression of the excited state along with its annihilation into ground states ($\chi^\ast \chi^\ast \to \chi \chi$) push the $\chi^\ast$ density to negligible values. The (diagonal) annihilation process, on the other hand, is strongly velocity-suppressed in the non-relativistic limit due to its $p$-wave nature~\cite{Duerr:2016tmh}. As a result, there are no relevant constraints on the annihilation cross section.

Finally, elastic scattering via the diagonal couplings can potentially give rise to sizeable DM self-interactions. The self-interaction cross section is approximately given by
\begin{equation}
    \sigma_\text{SIDM} \sim \frac{\alpha_\text{el}^2 m_\text{DM}^2}{m_{A'}^4} \; ,
\end{equation}
which is below current observational constraints for $m_\text{DM} = m_{A'}/3 > 10 \, \mathrm{MeV}$ and $\alpha_\text{el} < 0.5$.
}

{
\section{Collider searches}
\label{sec:colliders}

\begin{figure}
     \centering           
     \begin{subfigure}{0.42\textwidth}
        \centering
        
        \includegraphics[width=.99\linewidth]{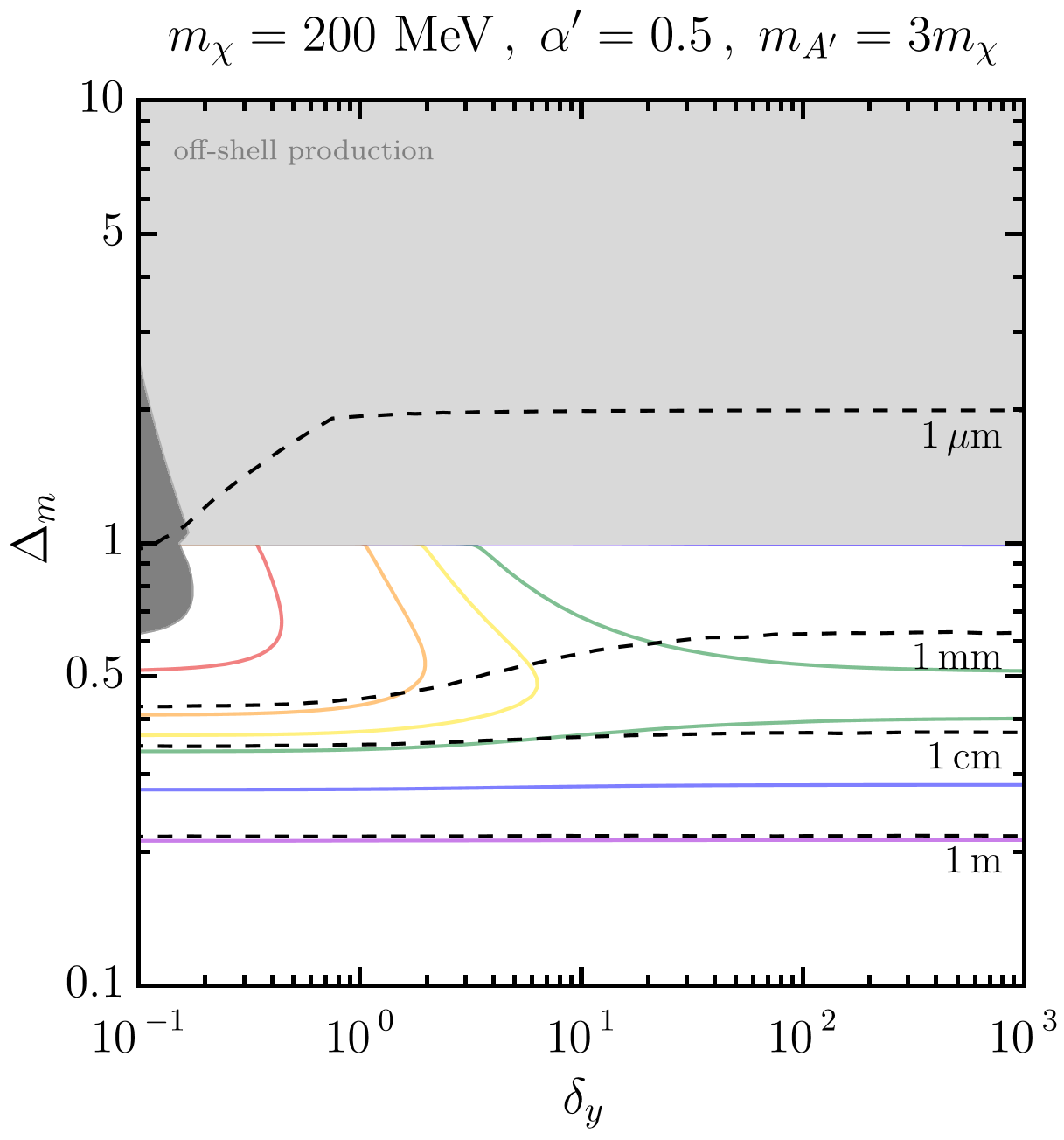}
     \end{subfigure}
     \hspace{0.05cm}
      \begin{subfigure}{0.49715\textwidth}
        \centering
        \includegraphics[width=.99\linewidth]{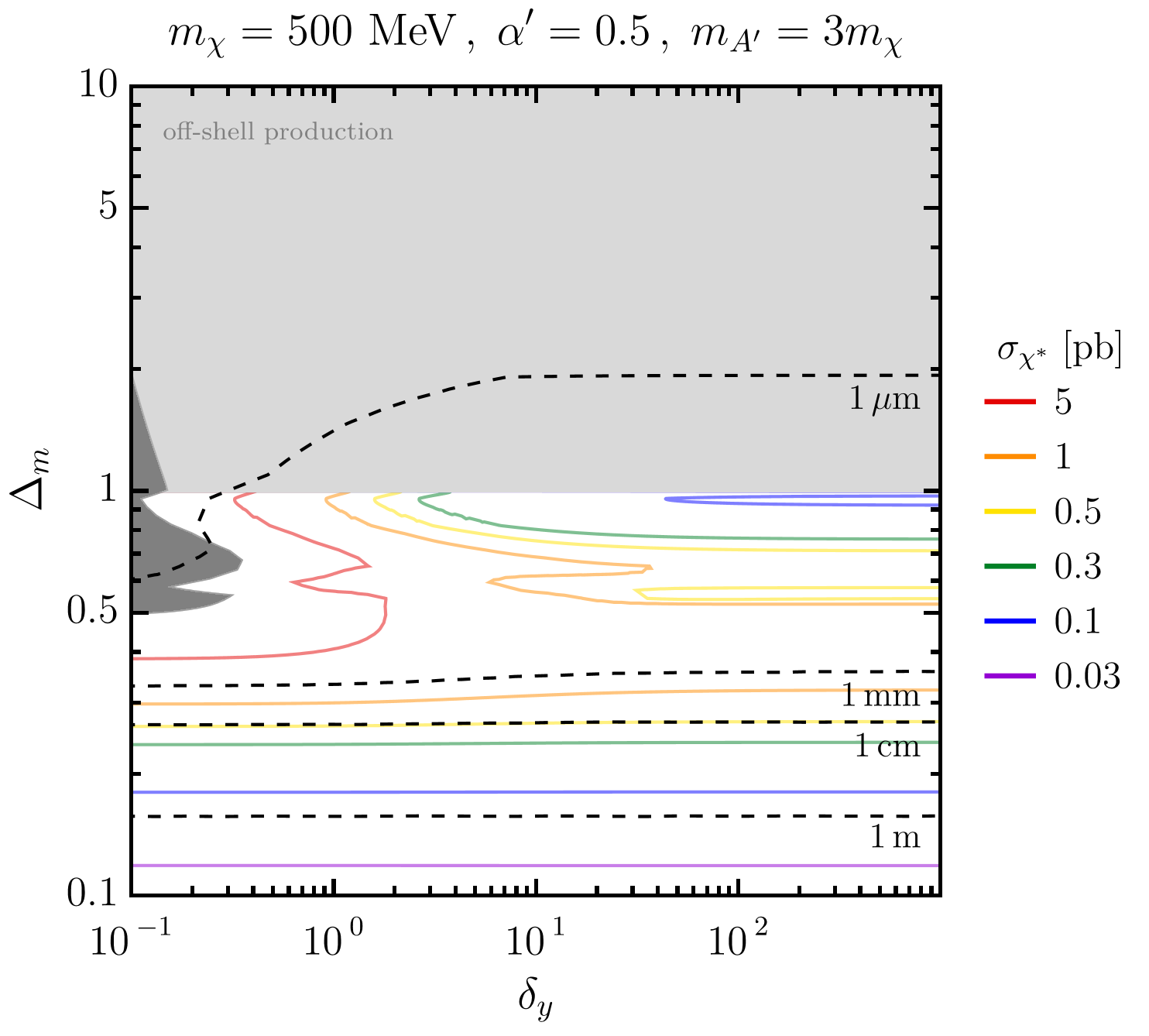}
     \end{subfigure}
        \caption{ Cross section for $\chi^\ast$ production as defined in eq.~\eqref{eq:sigma_chistar}  at an electron-positron collider with a center-of-mass energy of $E_{\text{CM}} = m_{\Upsilon(4S)} 
 = 10.58$~GeV for the thermal target $\epsilon_{\text{DM}}$.  Each panel stands for a different DM mass $m_{\chi}=(200,500)$~MeV. Dashed black lines show the proper decay length of $\chi^{\ast}$. In the light gray region the  $\chi^{\ast}$ production from $e^+ e^-$ annihilation would require an off-shell dark photon and is therefore strongly suppressed. Dark gray shaded regions are excluded by model-independent bounds on $\epsilon$.}
        \label{fig:collidersInfo1mass}
\end{figure}

The values of the kinetic mixing parameter $\epsilon$ that we obtained from the freeze-out calculation in section~\ref{sec:relic} are sufficiently large that we can hope to detect DM particles produced at current high-energy experiments, such as particle colliders and beam-dump experiments. As an illustration of possible signatures in this type of experiment, we present in \cref{fig:collidersInfo1mass} the cross section for $\chi^\ast$ production, defined as
\begin{equation}
\label{eq:sigma_chistar}
\sigma_{\chi^\ast} = \sigma(e^+ e^- \to A' \gamma) \times [\text{Br}(A' \to \chi \chi^{\ast}) + 2\text{Br}(A' \to \chi^{\ast} \chi^{\ast})] \; ,
\end{equation}
at the usual center-of-mass energies of $e^+e^-$ colliders, $E_{\text{CM}} = m_{\Upsilon(4S)} =10.58$~GeV. The production of the excited state is particularly interesting because for $\Delta_m \leq 2$ they decay into $\chi$ plus visible SM particles. 

The proper decay length of $\chi^{\ast}$ is also shown in \cref{fig:collidersInfo1mass} and should be compared to the typical size of central detectors in experiments like Belle II, which are in the range of 10\,cm--1\,m.  For $m_\chi = 200\,\mathrm{MeV}$ (left panel), we can conclude that possible $\chi^{\ast}$ signatures vary from decays outside the detector for $\Delta_m \lesssim 0.2$ to prompt decays for $\Delta_m \gtrsim 0.6$. In the intermediate range, $\Delta_m \approx 0.3$--0.5, the $\chi^\ast$ is expected to decay from a displaced vertex, which may be vetoed by conventional analyses but may be targeted with dedicated searches. For $m_\chi = 500\,\mathrm{MeV}$ (right panel) the decay width increases and therefore the lines of constant decay width shift to smaller values of $\Delta_m$.

Even if the $\chi^\ast$ has such a long decay length that it escapes from detectors close to the production point, its decays may still be observable at beam dump experiments. Moreover, one can also search for events with escaped particles using missing energy searches at active beam dumps, and one can search for the scattering of both excited states and ground states in downstream detectors. In the remainder of this section, we describe the details of these experiments and how we implement them in our analysis. Readers primarily interested in our results can directly skip to section~\ref{sec:results}.

\subsection{Proton beam dump experiments}

Proton beam dump experiments operate with a high-intensity beam of protons hitting a thick target. Hypothetical long-lived particles may be produced at such collisions and then decay inside a displaced decay volume. Past experiments have not observed such events, leading to exclusions in the parameter space.

For the model considered in this paper, the main signature would be decays of the excited state $\chi^{*}$. The parameter regions most interesting for niDM ($m_{\chi^\ast} \lesssim 1 \, \mathrm{GeV}$ and $\Delta_m \gtrsim 0.3$) correspond to relatively small lifetimes $c\tau_{\chi^{*}}\lesssim 1\text{ m}$, as shown in \cref{fig:collidersInfo1mass}. The number of events in this regime is exponentially suppressed by the decay probability, 
\begin{equation}
P_{\text{decay}} \approx \exp\left(-\frac{l \, m_{\chi^\ast}}{\tau_{\chi^{*}} |\mathbf{p}_{\chi^{\ast}}| }\right)\;, 
\end{equation}
where $l$ is the radial decay length. Therefore, the largest coupling that can be probed is determined by the ratio $ p_{\text{max}}/l_{\text{min}}$, where $p_{\text{max}}$ is the maximal possible momentum of $\chi^{*}$ particles decaying inside the decay volume, and $l_{\text{min}}$ is the distance from the $\chi^{*}$ production point to the beginning of the decay volume.

Calculating the event yield for $\chi^{*}$ decays requires several steps: calculating the flux of dark photons, decaying them into $\chi^{*}$, decaying $\chi^{*}$ inside the decay volume, propagating its decay products through the detector, and, finally, selecting the events which match all the experiment-specific criteria. To perform the calculations, we have updated \texttt{SensCalc}~\cite{Ovchynnikov:2023cry}, a \texttt{Mathematica}-based sensitivity evaluator.\footnote{The module is not public yet, but it may be provided on request.} By default, \texttt{SensCalc} calculates the event rate as a function of two parameters -- LLP mass and coupling to the SM particles, by averaging over decay products acceptance precomputed for fixed values of other parameters. This is not suitable for our purposes because of many different parameters controlling the event rate. We have therefore written a module that takes the input on the experiment geometry, the tabulated angle-energy distribution of dark photons, and the routine calculating the decay product acceptance, and then samples events with $\chi^{*}$ on-flight, similar to standard Monte-Carlo simulators. First, it samples 4-momenta of dark photons from the tabulated distribution and decays them into $\chi^{*}$ via the two decay processes $A'\to \chi\chi^{*}/ \chi^{*}\chi^{*}$. Next, it selects only those $\chi^{*}$s that point to the polar coverage of the decay volume, and samples azimuthal angles such that the trajectory

	 of $\chi^{*}$ intersects the decay volume. Then, it distributes the decay vertices inside the decay volume according to the value of the $c\tau_{\chi^{*}}$. Finally, for the given $\chi^{*}$s' kinematics, it generates their decays into $\chi$ and a pair of $e^+e^-,\mu^+\mu^-$ and $\pi^+\pi^-$,\footnote{As discussed in section~\ref{sec:decays}, heavier mesons play a sub-dominant role in the mass range under consideration.} propagates the decay products through the detector (accounting for the possible presence of a dipole magnet) and calculates the decay products' acceptance.

In \cref{app:beam-dump}, we briefly discuss the experiments that we consider, namely NuCal, CHARM, BEBC, and NA62 in dump mode, and we compare our predictions with ref.~\cite{Tsai:2019buq}. BEBC is considered for the first time in the context of (n)iDM and is found to give constraints similar to NuCal for small masses and lifetimes. NA62 is a currently running experiment. It has collected $1.4\cdot 10^{17}$ protons-on-target (PoT) in 2021~\cite{NA62:2023nhs} (without evidence for new physics), and aims to accumulate $10^{18}$ before LS3 (2025)~\cite{Antel:2023hkf}. Given the large beam energy and on-axis placement, we find that NA62 imposes the strongest constraints among the proton beam-dump experiments, although a dedicated event analysis performed by the NA62 collaboration would be required to confirm our findings (see \cref{app:beam-dump} for further discussion of NA62).

\subsection{NA64}

The fixed-target experiment NA64 at CERN employs a 100 GeV electron beam impinging on an active beam dump. The detector is essentially composed of a $\sim1$-meter long electromagnetic calorimeter (ECAL), corresponding to $\sim 40$ radiation lengths ($40X_0$), followed by a $\sim 6.5$-meters long hadronic calorimeter (HCAL), corresponding to $\sim 30$ nuclear interaction lengths, with a large high-efficiency veto counter (VETO) in between both calorimeters~\cite{Gninenko:2767779}. The NA64 collaboration searched for missing energy events with $E_{\text{missing}}>50$~GeV selected by requiring $E_{\text{ECAL}}<50$~GeV and $E_{\text{HCAL}}<1$~GeV with no activity in VETO, plus further requirements on the initial electrons which can be found in refs.~\cite{Banerjee:2019pds,NA64:2023wbi}. Such events could arise

	 from the production of high energy dark photons via electron-nucleus bremsstrahlung and via resonant annihilation of secondary positrons produced in electromagnetic showers with atomic electrons, followed by the subsequent decay of the dark photon into DM particles which escape the detector. 

In the niDM case, dark photons can also decay into excited states which, in principle, could decay inside the ECAL or within the HCAL. The former would decrease the total missing energy of the event (possibly leaving a signal in VETO from the electromagnetic shower of the daughter particles) and the latter would completely veto the event. As a result both possibilities decrease the sensitivity of the NA64 missing energy search to niDM, which is therefore most sensitive to the case where $\chi^\ast$ has a very long decay length. The corresponding effect in iDM models has been studied previously in the literature~\cite{Mongillo:2023hbs,Abdullahi:2023tyk} where proper simulations of the whole event were performed. Since the production rate and distribution of dark photons is identical for iDM and niDM, we can use these results to infer the relevant detector properties needed to analytically recast bounds from NA64 missing energy searches to the niDM case. Further details on this approach can be found in \cref{app:NA64details}.

It is worth noting that dedicated searches for semi-visible dark photon decays were also performed by NA64. Results for the iDM case~\cite{NA64:2021acr,Abdullahi:2023tyk} suggest that these searches would only probe parameter regions already excluded by proton beam dump experiments or by NA64 missing energy searches. Furthermore, new missing energy measurements with high energy muon beams at NA64 promise to test dark photon masses as large as 3~GeV, which potentially may probe the totality of the unexplored parameter space for sub-GeV masses~\cite{Gninenko:2653581}. Further studies are required to confirm these expectations.

\subsection{Electron-positron colliders}\label{sec:5p3missingEnergySearchesAtColliders}

Experiments at electron-positron colliders, such as PEP-II and SuperKEKB, operate at a fixed invariant mass and therefore have well-defined event kinematics and a much cleaner background environment than hadronic experiments (and, in particular, beam dumps). A particular search that is very sensitive to the production of dark photons and their subsequent decay into dark matter particles is the single-photon search $e^+e^- \to \gamma A'(\to \chi \chi)$, which looks for a single high energy photon in association with missing energy, i.e., $E_{\gamma} < E_{e^+}+E_{e^-}$. Events with additional particles in the final state (within the detector acceptance) are vetoed. This kind of search was performed by the BaBar collaboration~\cite{BaBar:2017tiz} and currently places the most stringent bounds on invisibly decaying dark photons in the mass range from 400~MeV to 8~GeV~\cite{Fabbrichesi:2020wbt,NA64:2023wbi}. Future single–photon searches at Belle~II are expected push these limits even further~\cite{Belle-II:2018jsg}. 

As for NA64, we expect these bounds to be relaxed in the context of niDM, since the dark photon decays can produce excited state particles, which may decay into SM particles inside the detector and lead to the event being vetoed. In order to analyse this effect, we have developed a \texttt{Mathematica}-based simulation for the full collision event.\footnote{Available at~\href{https://github.com/gdvgarcia/MissingEnergyEEcolliders}{https://github.com/gdvgarcia/MissingEnergyEEcolliders} .}  Events are generated at the centre-of-mass frame of the collision, then boosted to the laboratory frame.  First, we sample the dark photon production cross section $\sigma(e^+e^-\to\gamma A)$ and simulate the dark photon decays with the appropriate branching ratios into ground and excited states.\footnote{Strictly speaking, the process $e^+e^-\to\gamma \chi \chi^\ast$ should not be split into dark photon production and decay, since the width of the dark photon is non-negligible. However, this effect essentially only smears the photons and $\chi^{\ast}$ energies~\cite{Duerr:2019dmv}. From comparing our results to those found in the literature~\cite{Duerr:2019dmv}, we conclude that this effect is negligible within our precision goal.} If a $\chi^{\ast}$ particle is produced, we sample its decays length and decay modes. Cross sections, amplitudes and branching ratios are computed analytically at tree-level, using VMD for the $\chi^\ast \to \chi \pi^+ \pi^-$ decay mode, see section~\ref{sec:decays}. The 3-body decays into kaons, as well as 4-body and 5-body decays are expected to only give a small contribution in the mass range of interested and are therefore neglected. Following ref.~\cite{Duerr:2019dmv}, charged pions are treated as muons in the detector during the analysis of events.

Once events have been generated, we compute the single-photon constraints/sensitivity on the kinetic mixing $\epsilon_{\text{niDM}}$ by rescaling the results from searches for invisibly decaying dark photon~\cite{BaBar:2017tiz,Belle-II:2018jsg}  $\epsilon_{\text{mono-}\gamma}$ in the following way. We find the ratio $R_{\text{veto}}$ of the number of events which satisfy only the photon cuts but not the other selection requirements (i.e.\ the number of events which are vetoed) to the total number of events which satisfy the photon cuts (i.e.\ the number of events expected for a stable $\chi^\ast$). We then use $R_{\text{veto}}$ to compute the updated $\epsilon_{\text{exp}}^{\text{niDM}}$ according to \begin{equation}
    \epsilon_{\text{niDM}} = \epsilon_{\text{mono-}\gamma} /\sqrt{1-R_{\text{veto}}} \, .
\end{equation} The above formula defines $\epsilon_{\text{niDM}}$ implicitly since $R_{\text{veto}}$ depends itself on $\epsilon_{\text{niDM}}$. 

A description of the conditions and selection criteria for both BaBar and Belle II can be found in~\cref{app:colliders}. In this appendix we also compare the results of our simulation to Belle II sensitivities and BaBar bounds for iDM found in ref.~\cite{Duerr:2019dmv} finding a good agreement. 

\subsection{Searches for dark matter scattering}

If dark photons can be produced in particle collisions, we expect a sizeable flux of DM particles, both ground and excited states, passing through various detectors at high energy facilities. These particles can scatter relativistically on the active material of these experiments, leaving observable signals. Measurements performed at LSND~\cite{deNiverville:2011it,LSND:2001akn}, E137~\cite{Batell:2014mga,Bjorken:1988as} and MiniBooNE~\cite{MiniBooNE:2017nqe} place strong constraints on light DM with masses below a few hundred MeV. We adopt the bounds from ref.~\cite{Berlin:2018pwi}, which were obtained by rescaling the published results by the appropriate choice of $\alpha'$. They apply to our model independently of our choices of ($\delta_y,\Delta_m$) -- at least in the parameter range we focus on -- since $\chi^{(\ast)}$ are produced with relativistic velocities and energies much higher than the mass splitting between the two states.

}

{\section{Results}
\label{sec:results}

Having discussed the different experiments that can probe the niDM model, we are now in the position to determine the allowed regions of parameter space. For this purpose, let us first consider two particular DM masses, 200 and 500 MeV, and explore in a quantitative manner the two-dimensional plane of the dark left-right Yukawa asymmetry~$\delta_y$ and the relative mass splitting between dark sector fermions~$\Delta_m$ previously presented qualitatively in \cref{fig:spectrum}. The masses were chosen in order to illustrate the potential of different collider searches and their complementarity. 

In \cref{fig:3DcolliderExclusionPlots}, we present the various constraints on the niDM parameters. At each point in the $(\delta_y, \Delta_m)$ parameter plane we determine the value of the kinetic mixing parameter needed to reproduce the observed DM relic abundance and verify if the corresponding value is experimental excluded. Displayed in blue are the strongest constraints from proton beam dump experiments obtained by searching for $\chi^{\ast}$ decays inside the NA62 experiment~\cite{NA62:2017rwk}, in yellow from electron beam dumps via missing energy searches of high-energy electron collisions in the active target of NA64~\cite{NA64:2023wbi} and in green from electron-positron collisions using single-photon measurements ($e^+ e^- \to \gamma + \text{invisible}$) at BaBar~\cite{BaBar:2017tiz}. Also shown in gray are model-independent bounds on the kinetic mixing $\epsilon$  from electroweak precision measurements at LEP calculated in ref.~\cite{Hook:2010tw} and bounds from the scattering of $\chi^{(\ast)}$ particles at far detectors obtained from measurements performed at LSND, E137, and MiniBooNE in ref.~\cite{Berlin:2018pwi}. The light gray band indicates when the co-annihilation channel becomes resonant due $m_{\chi}+m_{\chi^\ast} = m_{A'} \pm 1\%$. Black lines indicate a constant ratio of  diagonal to off-diagonal couplings.

\begin{figure}
        \centering
     \begin{subfigure}{0.44\textwidth}
        \centering
        \includegraphics[width=.99\linewidth]{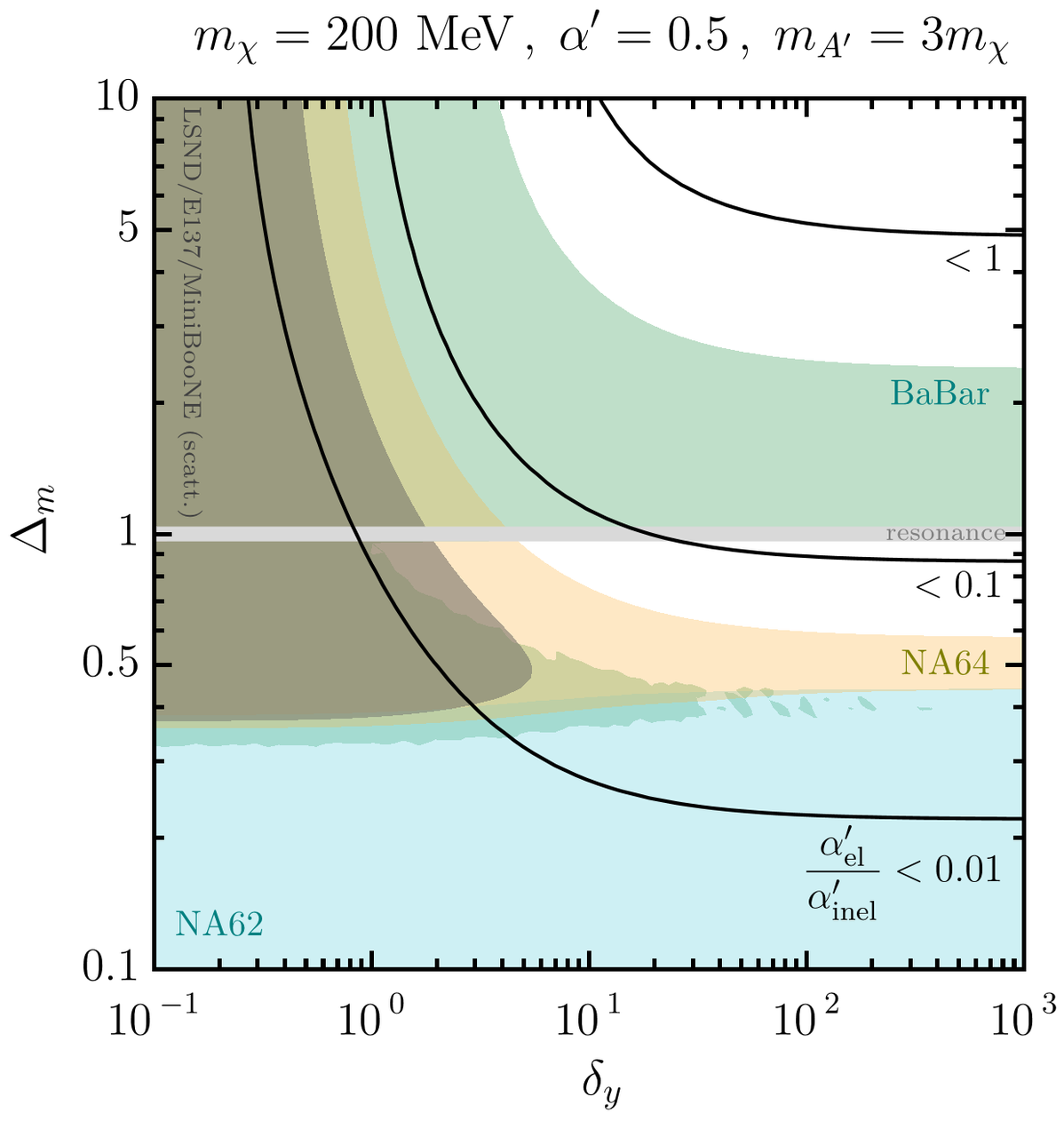}
     \end{subfigure}
     \hspace{0.05cm}
      \begin{subfigure}{0.44\textwidth}
        \centering
        \includegraphics[width=.99\linewidth]{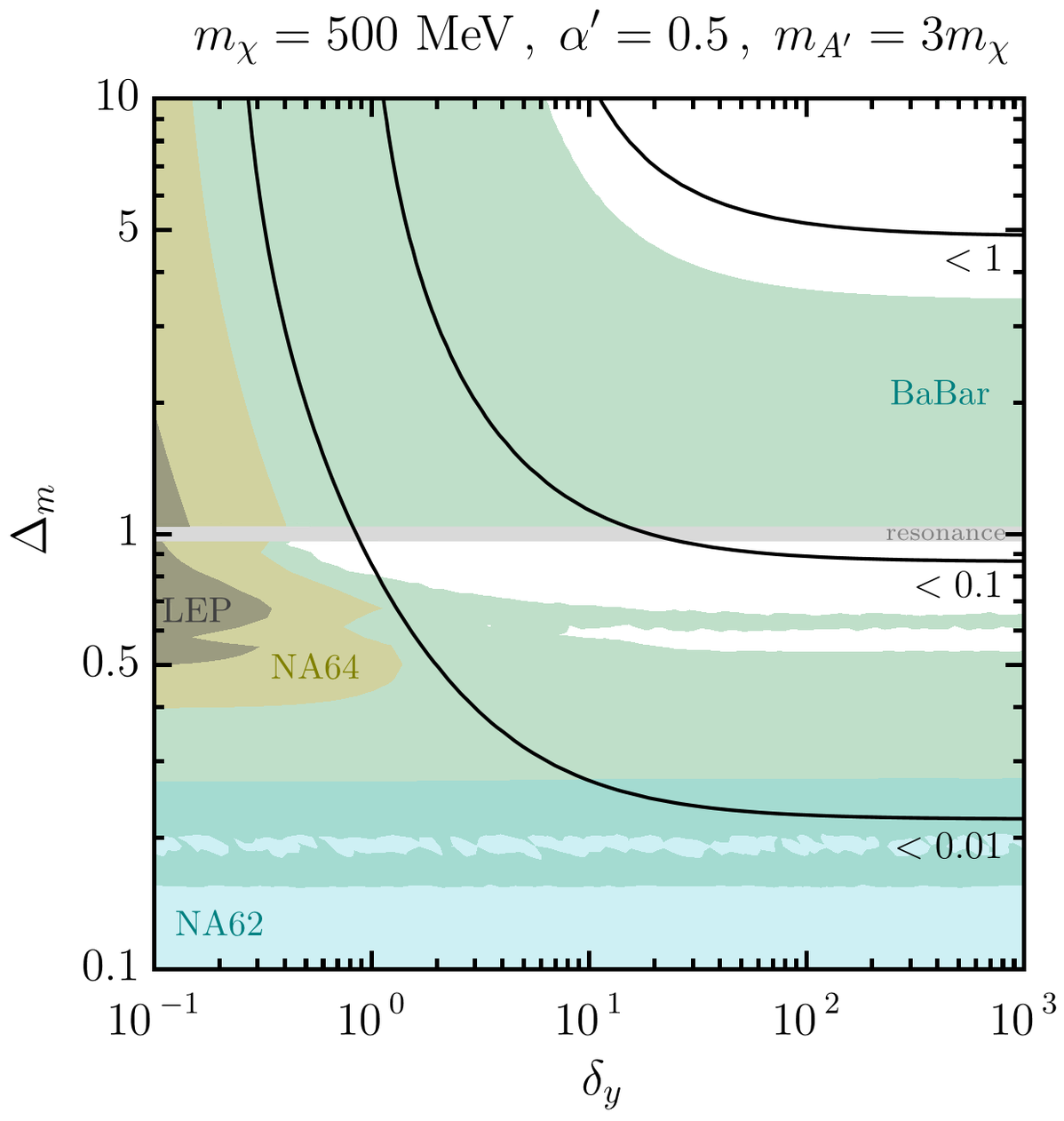}
     \end{subfigure}
        \caption{Region in the $(\delta_y,\Delta_m)$ parameter plane where the kinetic mixing required to reproduce the observed DM relic abundance is experimentally excluded. Black lines indicate the ratio of elastic (diagonal) $\alpha'_{\text{el}}$ to inelastic (off-diagonal) $\alpha'_{\text{inel}}$ couplings.
        }
        \label{fig:3DcolliderExclusionPlots}
\end{figure}

The first conclusion to draw from \cref{fig:3DcolliderExclusionPlots} is that the traditional regime of inelastic DM, where the diagonal interactions are negligible corresponding to $\delta_y \ll 1$ or $\Delta_m \ll 1$ is fully excluded for both DM masses and $(\delta_y,\Delta_m)$ ranges that we considered. On the other hand, in the Majorana DM regime ($\Delta_m \gg 1$ and $\delta_y \gg 1$), where the heavier state decouples and diagonal interactions dominate, all experimental constraints can be satisfied. As $\Delta_m$ approaches 1 from above, the constraint from BaBar gets stronger as a result of the larger kinetic mixing implied by the relic density requirement.

For $\Delta_m < 1$, however, the constraints from BaBar are significantly suppressed, because the decay $A' \to \chi \chi^\ast$ becomes kinematically allowed and the subsequent $\chi^\ast$ decays lead to missing-energy events being vetoed in the analysis. This opens up a sizeable allowed parameter region with $\Delta_m \sim 0.6$--0.9 and $\delta_y \gtrsim 1$. This is exactly the niDM regime, where diagonal interactions are relevant for the relic density calculation and  off-diagonal   signatures dominate at laboratory experiments.

The allowed regions in \cref{fig:3DcolliderExclusionPlots} are expected to be probed in the near future by a variety of currently running experiments. These include the NA62 experiment in proton beam dump mode~\cite{NA62:2017rwk} with its plans of collecting $\sim10^{18}$~PoT worth of data by 2025~\cite{Antel:2023hkf} which, however, are expected to provide minor upgrades from their previous results last year~\cite{NA62:2023nhs} due a logarithmic dependence of the kinetic mixing bound on the integrated intensity. More remarkable is the Belle II experiment~\cite{Belle-II:2018jsg} which has collected already more than 424~fb$^{-1}$ of collision data~\cite{Tenchini:2023wow} and, with only a luminosity of 20~fb$^{-1}$, could significantly improve the single-photon bound from BaBar covering all unexplored regions in the 2D plane of both panels in \cref{fig:3DcolliderExclusionPlots}. Additionally, another direction for significant progress is expected to come from future data releases of the NA64 collaboration which continues to collect more electrons-on-target (EoT) with a final goal of $10^{13}$~EoT within the next decade~\cite{Na64SPSC}. 

\begin{figure}
     \centering
     \begin{subfigure}{0.44\textwidth}
        \centering
        \includegraphics[width=.99\linewidth]{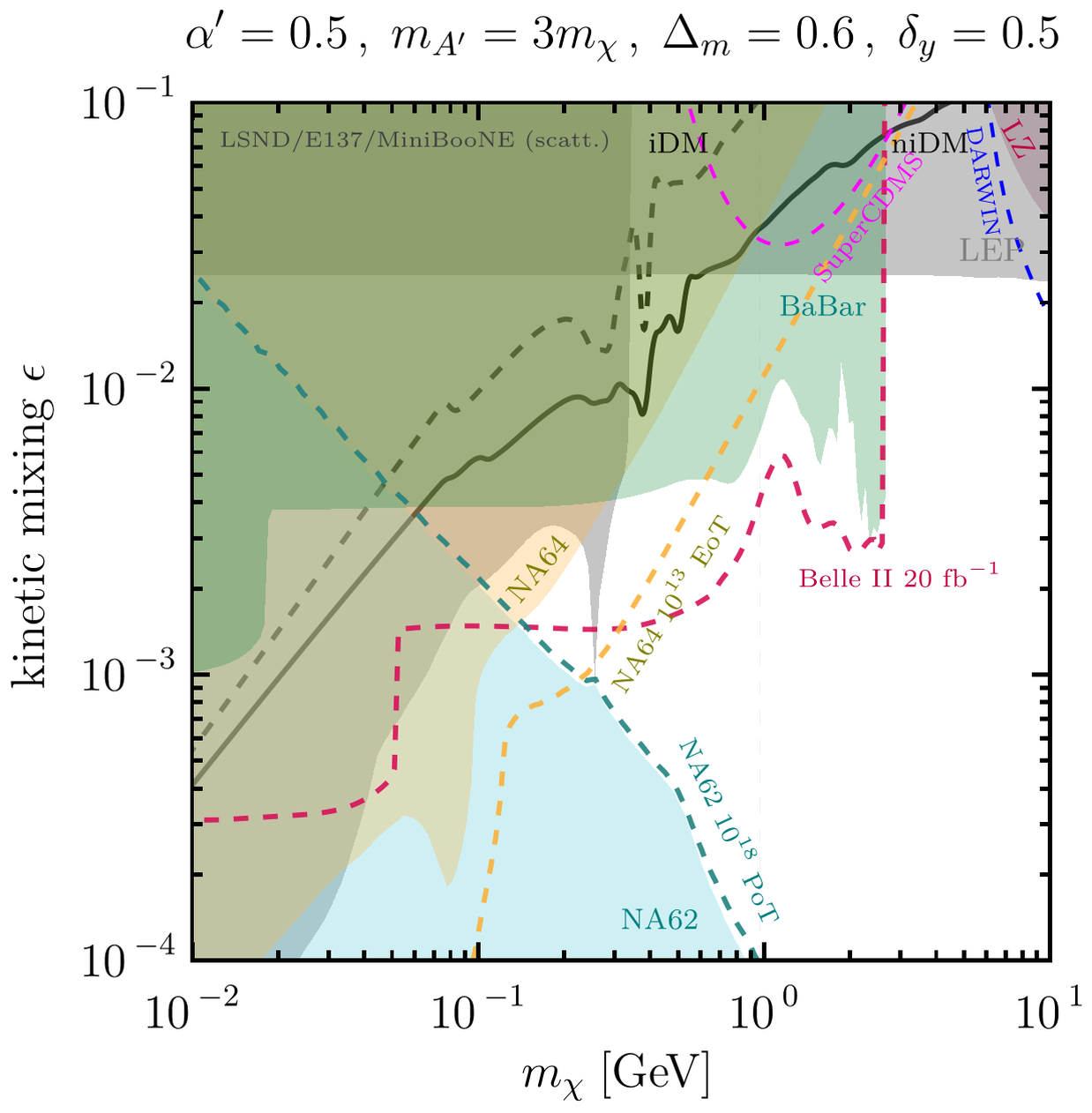}
     \end{subfigure} \hspace{0.05cm}
      \begin{subfigure}{0.44\textwidth}
        \centering
        \includegraphics[width=.99\linewidth]{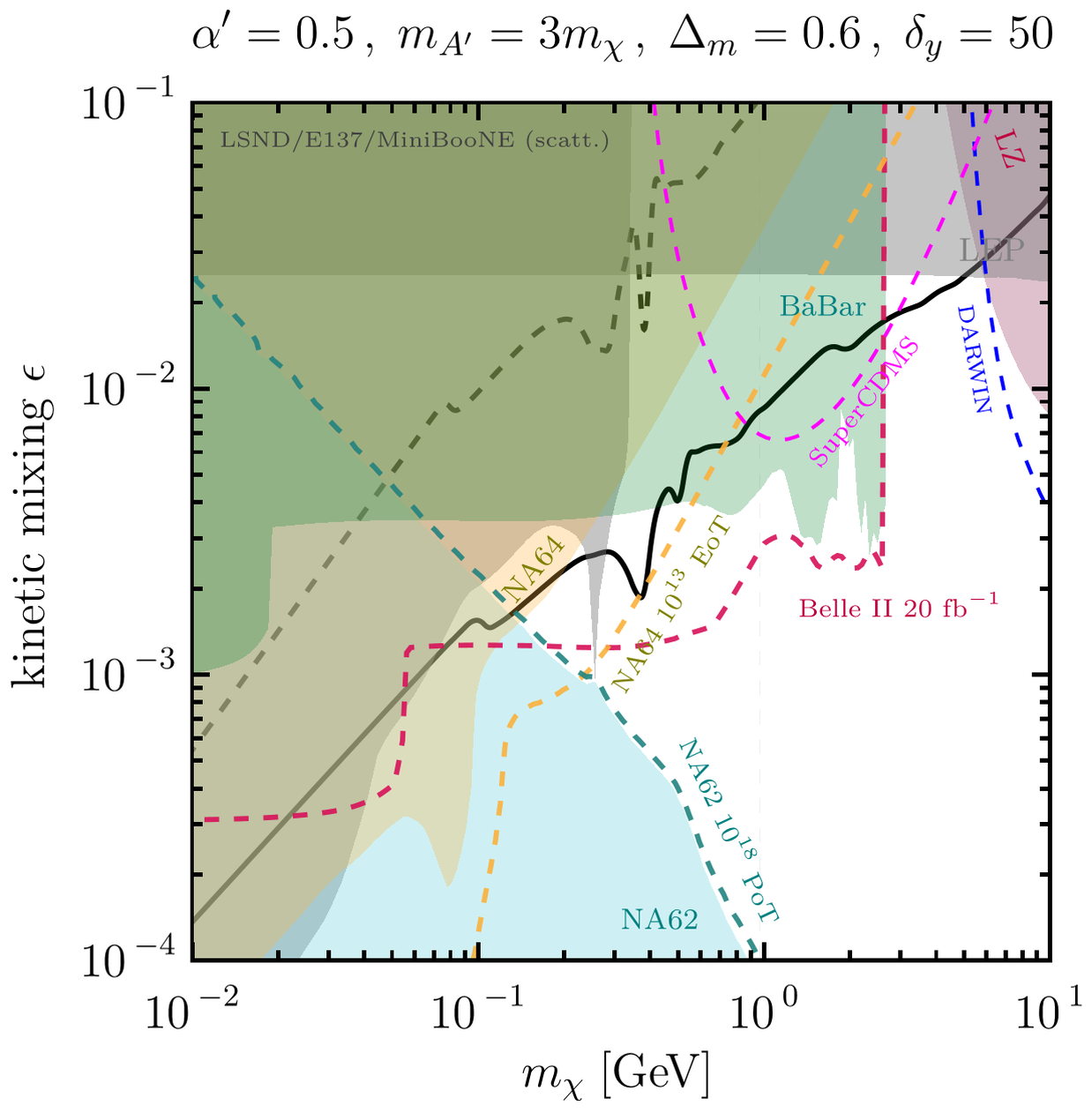}
     \end{subfigure}
     \begin{subfigure}{0.44\textwidth}
        \centering
        \includegraphics[width=.99\linewidth]{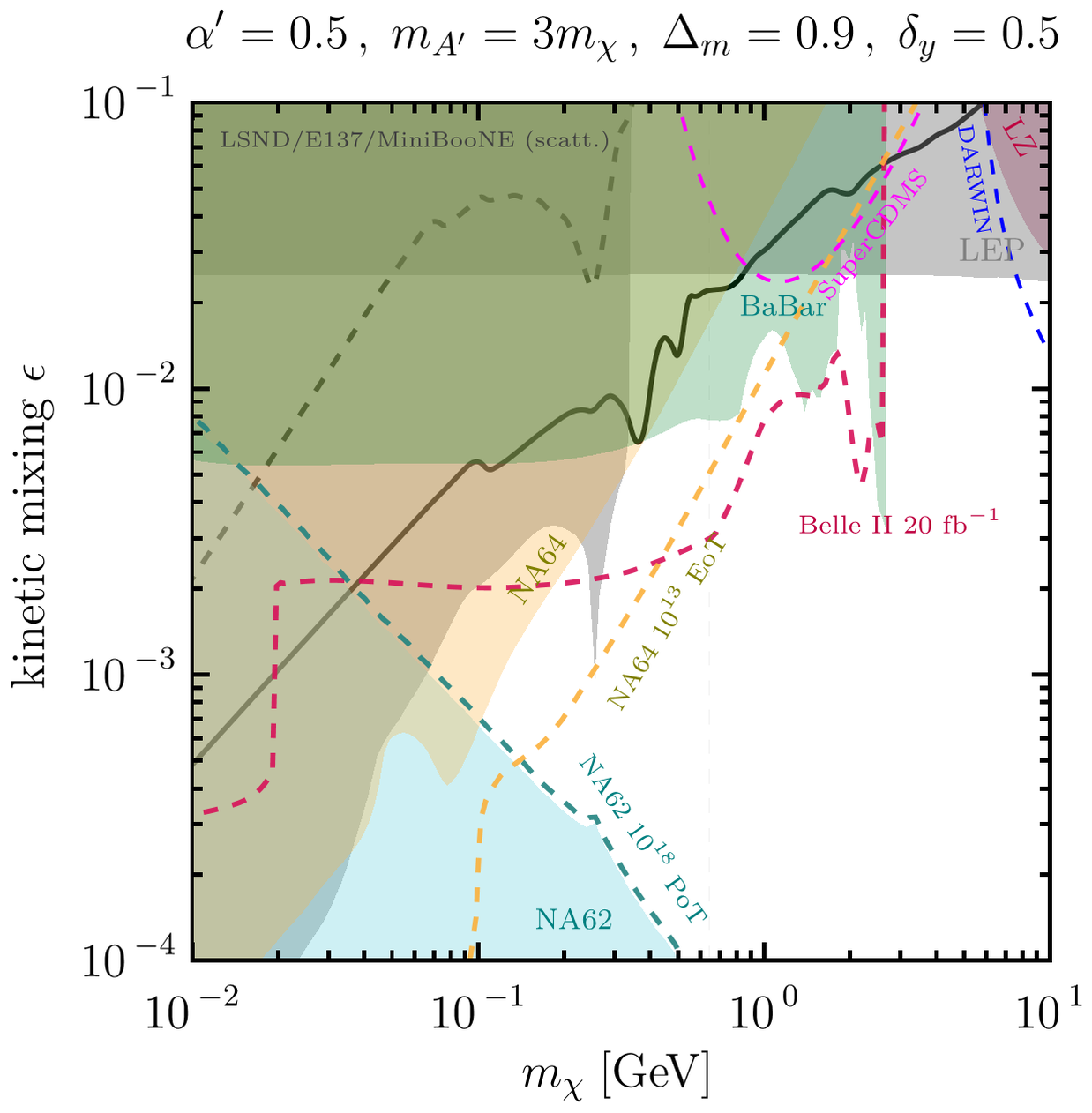}
     \end{subfigure}
     \hspace{0.05cm}
      \begin{subfigure}{0.44\textwidth}
        \centering
        \includegraphics[width=.99\linewidth]{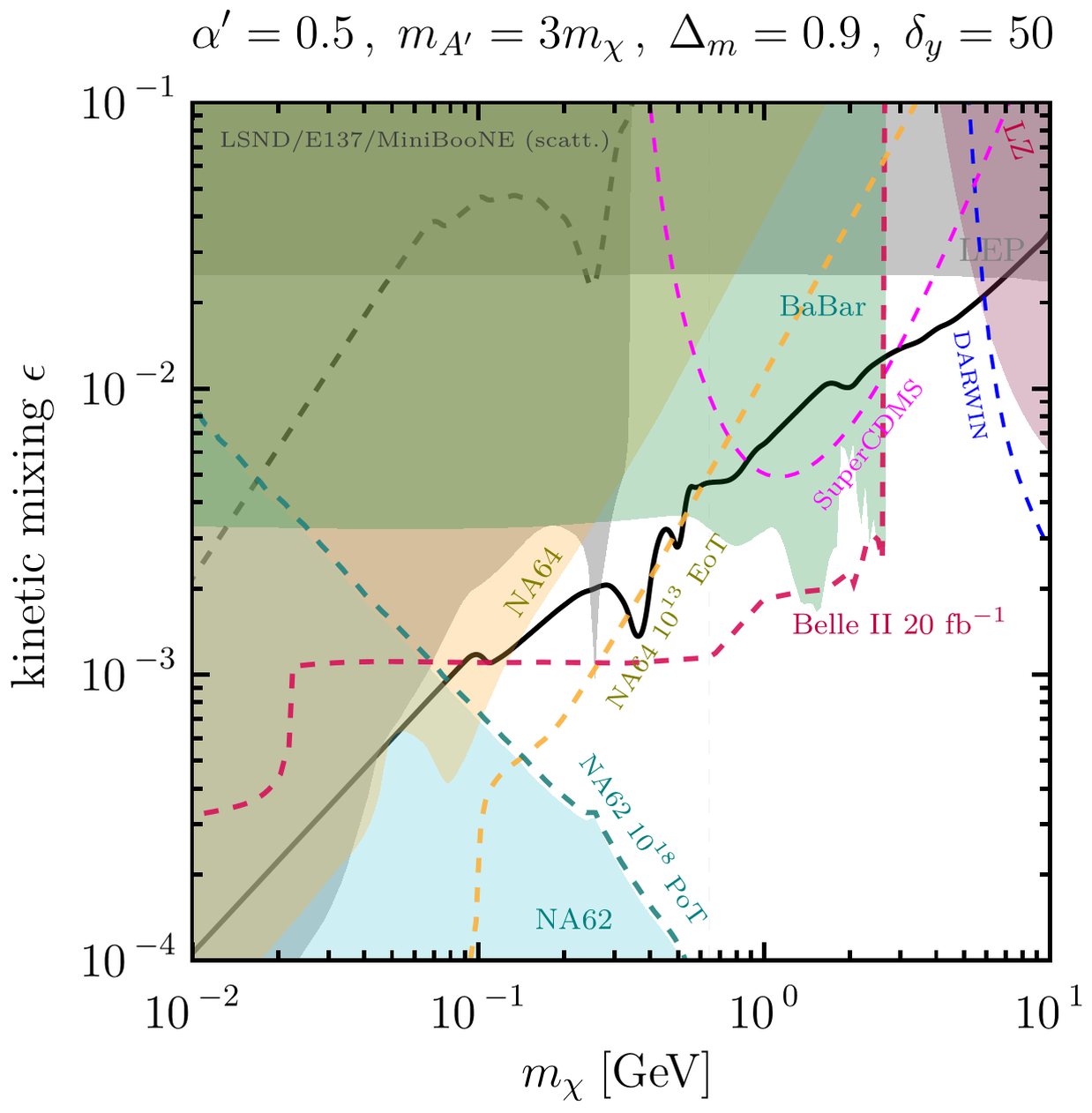}
     \end{subfigure}
        \caption{Current bounds and projected sensitivity to the kinetic mixing $\epsilon$ as a function of the DM mass in the niDM model. Solid black curves indicate the DM relic density curve for the given choice of $(\Delta_m,\delta_y)$ of each panel while, dashed ones indicate the relic density curve for iDM ($\delta_y = 0$) with the same mass splitting $\Delta_m$. 
        }
        \label{fig:finalPlotEpsXmassWithConstraints}
\end{figure}

The impact of these near-future improvements can be more clearly visualized looking at the plane of kinetic mixing $\epsilon$ versus DM mass presented in \cref{fig:finalPlotEpsXmassWithConstraints}. Different rows correspond to different values of $\Delta_m$, while different columns correspond to different values of $\delta_y$, chosen to lie in the most interesting parameter region identified above. The value of the kinetic mixing parameter that reproduces the observed DM relic abundances is shown in black and can be compared to the black dashed curve that one would obtain for purely inelastic DM (i.e.\ setting $\delta_y=0$ but keeping the same $\Delta_m$). The other dashed lines correspond to the projected sensitivites of the various currently running experiments. We find that collecting $10^{18}$~PoT with NA62 will, as expected, only increase the lower bound on $\epsilon$ by a few per cent, while NA64 will improve the upper bound on $\epsilon$ by more than a factor of 3. The most decisive improvement, however, will come from Belle II, which will probe the relic density line for DM masses below $\sim 3$~GeV for all parameter choices we have considered. Furthermore, new missing energy measurements with high energy muon beams at NA64 may test dark photon masses as large as 3~GeV, with a potentially similar sensitivity reach compared to the Belle II mono-photon analysis~\cite{Gninenko:2653581} (not shown in \cref{fig:finalPlotEpsXmassWithConstraints}).

In addition to bounds from collider experiments, we also show in \cref{fig:finalPlotEpsXmassWithConstraints} exclusion curves for the direct detection experiment LZ~\cite{LZCollaboration:2024lux} which probes large DM mass values $m_{\chi} \gtrsim 7$~GeV and is therefore highly complementary to the bounds from particle accelerators. We also show projections for future direct detection experiments, namely the far-future DARWIN experiment~\cite{Schumann:2015cpa}, which targets the same mass region as LZ, and the near-future SuperCDMS experiment~\cite{SuperCDMS:2016wui}, which is sensitive to smaller DM masses. We find that for $\delta_y = 50$ there is an intriguing gap in coverage between these experiments for $m_\chi \sim 5\,\mathrm{GeV}$, where direct detection experiments face a challenging background from solar $^8$B neutrinos~\cite{Billard:2013qya}. Exploring possible experimental opportunities to cover this gap is an exciting direction for future research.
}

{
\section{Conclusions}
\label{sec:conclu}

In the present work we have studied the consequences of relaxing the assumption of parity conservation in the dark sector of inelastic Dark Matter (iDM) models, a symmetry which is often assumed to be conserved (or only weakly broken) in the literature. As we discussed, this assumption is not necessary. Its breaking does not spoil any of the attractive properties of iDM but instead open up new viable parameter space. We call this model ``not-so-inelastic Dark Matter'' (niDM), because, in this generic extension of iDM, elastic (diagonal) interactions $\chi X \to \chi X$ are allowed in addition to inelastic (off-diagonal) ones $\chi X \to \chi^\ast X$. Just like in iDM models, the excited DM state $\chi^\ast$ can offer interesting decay signatures at particle physics experiments.

We have introduced a UV-complete niDM model for fermionic DM particles, based on an abelian gauge symmetry in the dark sector. The coupling to the Standard Model (SM) proceeds via kinetic mixing of the dark vector boson $A'$ with the SM photon. The model provides a natural interpolation between conventional iDM models and simple Majorana DM where the excited state decouples. The relevant parameters are the relative mass splitting of the two dark fermions, $\Delta_m$, and a parity breaking asymmetry parameter, $\delta_y$, where $\delta_y\to 0$ corresponds to the iDM case. The niDM regime is characterised by $\Delta_m \sim 1$ and $\delta_y \gtrsim 1$. While the iDM limit of the model is strongly constrained (mostly excluded), we found that allowing for non-zero parity asymmetry $\delta_y$ opens up new parameter space, where the correct relic DM density can be obtained by thermal freeze-out without being in conflict with existing bounds. For our benchmark model, we identified two windows for the DM mass namely $m_\chi \simeq 400$~MeV and $m_\chi \simeq 5$~GeV.  

We studied in detail the phenomenology of the model in the relevant regions of parameter space. For the allowed region around $m_\chi \sim 5$~GeV, direct detection experiments such as LZ, or the future SuperCDMS or DARWIN projects are potentially sensitive, but the relic density curve remains difficult to test as it is obscured by the solar neutrino fog. Identifying new experimental opportunities to cover this gap is an exciting goal for future research.

In contrast, for sub-GeV masses, signatures in high energy experiments offer promising tests of the model, as the production and subsequent decay of the excited DM state leads to interesting phenomenology and new signatures. We performed a detailed simulation of beam dump experiments (NuCal, CHARM, BEBC, NA62 in beam dump mode) and electron-positron colliders (BaBar, Belle II) along with an analytical description of NA64 missing energy searches, and recasted their results within the niDM model. We found that large regions of parameter space were excluded by a combination of all available constraints, leaving only the above mentioned window of DM masses around 400~MeV for suitable values of $\Delta_m$ and $\delta_y$. Near-future analyses of NA64 and Belle-II (mono-gamma searches) will cover the entire parameter space we studied for sub-GeV niDM, where the model can provide the full DM abundance. Furthermore, new missing energy measurements with high energy muon beams at NA64 may provide competitive sensitivity to the allowed parameter space. Hence, in this mass range, currently running experiments will fully explore the potential of not-so-inelastic Dark Matter as a suitable dark matter candidate, pushing the elasticity of the model to its limits with a possible break(through) right here at our hands.
}

{\acknowledgments{The authors thank Nassim Bozorgnia, Babette D\"{o}brich, Torber Ferber, Saniya Heeba, Laura Molina Bueno and Javier Reynoso-Cordova for discussions and Yu-Dai Tsai for answering questions on ref.~\cite{Tsai:2019buq}. The authors also thank Nassim Bozorgnia and Javier Reynoso-Cordova for providing the DM distributions used in ref.~\cite{Reynoso-Cordova:2024xqz}.   GG thanks the Doctoral School  ``Karlsruhe School of Elementary and Astroparticle Physics: Science and Technology (KSETA)” for financial support through the GSSP program of the German Academic Exchange Service (DAAD).
This work has received support by the European Union’s Framework Programme for Research and Innovation Horizon 2020 under grant H2020-MSCA-ITN-2019/860881-HIDDeN.
FK acknowledges funding by the Deutsche Forschungsgemeinschaft (DFG) through the Emmy Noether Grant No.\ KA 4662/1-2 and Grant No.\ 396021762 -- TRR 257.}
}

\appendix

{\section{niDM in full generality}
\label{app:general-model}
\Cref{eq:NewPhysicsLagrangian} constitutes the most general new-physics Lagrangian containing the new dark sector particles that we consider. In our analysis, this Lagrangian was simplified with three assumptions, namely $\phi_d=0$ and $m_d \geq 2w\sqrt{y_R y_L}$ along with $m_{A'} \ll m_Z$. Moreover, we have required a heavy dark Higgs field with a weak Higgs portal in order to neglect the scalar sector of our model. In this appendix we revisit these assumptions. We will introduce the most general formulae for interactions in the dark sector after the unitary transformation which diagonalizes the fermionic mass matrix keeping all masses positives. 

The dark fermionic Lagrangian can generally be written as \begin{equation}
    \mathcal{L}_{\chi} = \dfrac{1}{2} \Bar{\chi}_i  (i \slashed{\partial}  - m_i) \chi_i +   \dfrac{1}{2} g_{\chi} A'^{\mu} \, \Bar{\chi}_i  ( i \alpha_{ij} + \beta_{ij}  \gamma^{5} ) \gamma_{\mu} \chi_j - \dfrac{1}{2} y_L s   \,\Bar{\chi}_i  ( \hat{\alpha}_{ij} + i \hat{\beta}_{ij}  \gamma^{5} ) \chi_j
\end{equation} where the masses of the Majorana fermions are given by \begin{align}
   & m_{\chi^{\ast}}^2 = m_d^2+ 2 w^2 (y_L^2 + y_R^2) + 2 w D && \text{and} &&  m_{\chi}^2 =  m_d^2+ 2 w^2 (y_L^2 + y_R^2) - 2 w D \, , 
\end{align} with\footnote{ The particular case of $\kappa=0$ (i.e., $\cos \phi_d =0 $ and $y_R=y_L$) leads to no mass splitting between the Majorana states and is therefore not considered further in this work.} \begin{align}
    D^2 = w^2(y_R^2-y_L^2)^2 + m_d^2 \abs{\kappa}^2 && \text{and} && \kappa = y_L e^{i \phi_d} + y_R e^{-i \phi_d}\, .
\end{align} We adopt $D\geq0$, since we define $\chi^{\ast}$ to be the heavier state without loss of generality. The interaction coefficients are given in \cref{tab:InteractionCoefficientsVector,tab:InteractionCoefficientsScalar} where the lower-script ``$\ast$'' denotes the excited state index and ``$\_$'' the ground state one. The complex phase $\phi$ is defined as \begin{align}
   e^{i\phi} =  \dfrac{\kappa}{\abs{\kappa}}\, .
\end{align} The phases $\gamma_{i}$ are given by \begin{align}
   e^{i \gamma_{\ast}} = \dfrac{\sigma_{\ast}}{\abs{\sigma_{\ast}}}    && \text{and} && e^{i \gamma_{\_}} =\dfrac{\sigma_{\_}}{\abs{\sigma_{\_}}}  \, ,
\end{align} where \begin{equation}
    \sigma_{\ast}  =\dfrac{1}{2}\dfrac{w(y_L^2 + 2 
 y_L y_R  e^{-2i\phi_d}  +y_R^2)+D}{ \kappa \,e^{-i\phi_d}}  \end{equation} and \begin{equation} \sigma_{\_} = \dfrac{1}{2}\dfrac{w(y_L^2 + 2 
 y_L y_R  e^{-2i\phi_d}  +y_R^2)-D}{\kappa^\ast \,e^{-i\phi_d}} \, .
\end{equation} Finally, the mixing angle $\theta$ is given by \begin{equation}
    \tan{\theta} =  \dfrac{w(y_R^2-y_L^2)+D}{m_d\abs{\kappa}} \,.
\end{equation}

\begin{table}
\centering
\caption{\label{tab:InteractionCoefficientsVector} Coefficients for interactions between dark fermions and dark photons.  } 
\begin{tabular}{cccc}

\hline
$\alpha_{\ast\ast} = 0 $& $\beta_{\ast\ast} = - \cos{2\theta} $\\

$\alpha_{\ast\_} =   \sin{( (2 \phi + \gamma_{\ast} - \gamma_{\_} )/2)} \sin{2\theta} $& $\beta_{\ast\_} =  \cos{( (2 \phi + \gamma_{\ast} - \gamma_{\_} )/2)} \sin{2\theta} $ \\

$\alpha_{\_\ast} = - \alpha_{\ast\_} $& $\beta_{\_\ast} = \beta_{\ast\_}  $ \\

$\alpha_{\_\_} = 0 $& $\beta_{\_\_} = \cos{2\theta} $ \\
\hline
\end{tabular}
\end{table}

\begin{table}
\centering
\caption{\label{tab:InteractionCoefficientsScalar} Coefficients for interactions between dark fermions and dark Higgs.  } 
\begin{tabular}{cccc}

\hline

 \multicolumn{2}{c}{$ \hat{\alpha}_{\ast\ast} = 2 (\delta_y + 1) \cos{(2\phi +\gamma_\ast)} 
  \sin^2 \theta+ 2 \cos{\gamma_\ast} \cos^2 \theta $}\\
  
  \multicolumn{2}{c}{$ \hat{\beta}_{\ast\ast} =  2(\delta_y + 1) \,\sin{(2\phi +\gamma_\ast)} 
  \sin^2 \theta+2 \,\sin{\gamma_\ast} \cos^2 \theta$} \\ 
  
  \multicolumn{2}{c}{$ \hat{\alpha}_{\ast\_} = \sin{2\theta}\big[(\delta_y +1) \cos{((2 \phi + \gamma_\ast + \gamma_\_)/2)} - \cos{((2 \phi - \gamma_\ast - \gamma_\_)/2)}  \big]$ } \\
  
  \multicolumn{2}{c}{$ \hat{\beta}_{\ast\_} =  \sin{2\theta}\big[ (\delta_y +1) \,\sin{((2 \phi + \gamma_\ast + \gamma_\_)/2)} + \,\sin{((2 \phi - \gamma_\ast - \gamma_\_)/2)}  \big]$} \\
  
   \multicolumn{2}{c}{ $ \hat{\alpha}_{\_\ast} = \hat{\alpha}_{\ast\_}$ \;\;\;\;\; \;\;\;\;\; \;\;\;\;\; \;\;\;\;\; \;\;\;\;\; \;\;\;\;\; $ \hat{\beta}_{\_\ast} = \hat{\beta}_{\ast \_}$}\\
   
  \multicolumn{2}{c}{ $ \hat{\alpha}_{\_\_} =  2(\delta_y + 1) \cos{\gamma_\_}  
  \cos^2 \theta  + 2\cos{(2\phi -\gamma_\_)} \sin^2 \theta$ }\\
  
  \multicolumn{2}{c}{$ \hat{\beta}_{\_\_} = 2(\delta_y + 1) \,\sin{\gamma_\_}  
  \cos^2 \theta  -2 \,\sin{(2\phi -\gamma_\_)} \sin^2 \theta $ }\\
  
\hline
\end{tabular}
\end{table}

Compared to the simpler case with $\phi_d = 0$, the only modification (as long as the interactions of the dark Higgs boson are negligible) is that the inelastic transition can also proceed via an axial-vector coupling rather than only a vector one. As a result, the co-annihilation contribution to the relic density is slightly modified.

Regarding the mediators, one can find the formulas related to dark Higgs mixing with the SM and its diagonalization in ref.~\cite{Duerr:2016tmh} while the general $U'(1)$ kinetic mixing diagonalization, in leading order in $\epsilon$, is simply realized by the field redefinitions \begin{align} 
    & A^{\mu} \to A^{\mu} = \hat{A}^{\mu} - \epsilon \hat{A}'^{\mu} \\ 
    & Z^{\mu} \to Z^{\mu} = \hat{Z}^{\mu} + \dfrac{m_{A'}^2}{m_{A'}^2 - m_{Z}^2} \epsilon \tan{\theta_w} \hat{A}'^{\mu}\\
    & A'^{\mu} \to A'^{\mu} = \hat{A}'^{\mu} + \dfrac{m_{Z}^2}{m_{Z}^2 - m_{A'}^2} \epsilon \tan{\theta_w} \hat{Z}^{\mu}
\end{align}  where the fields with hats correspond to the physical states, and we have assumed that $|m_Z / (m_Z - m_{A'})| < 1$. The transformations above show that the $A'$ boson can interact with SM particles through both mixing with the photon and with the $Z$ boson. Also, we see that the $Z$ boson obtains direct couplings to the dark sector particles, which is suppressed by $\epsilon \tan{\theta_w} m_{Z}^2/(m_{Z}^2-m_{A'}^2)$.
}

{\section{Impact of dark matter halo modeling on direct detection limits}
\label{app:DDupdates}

\begin{figure}
    \centering
    \includegraphics[width=0.5\textwidth]{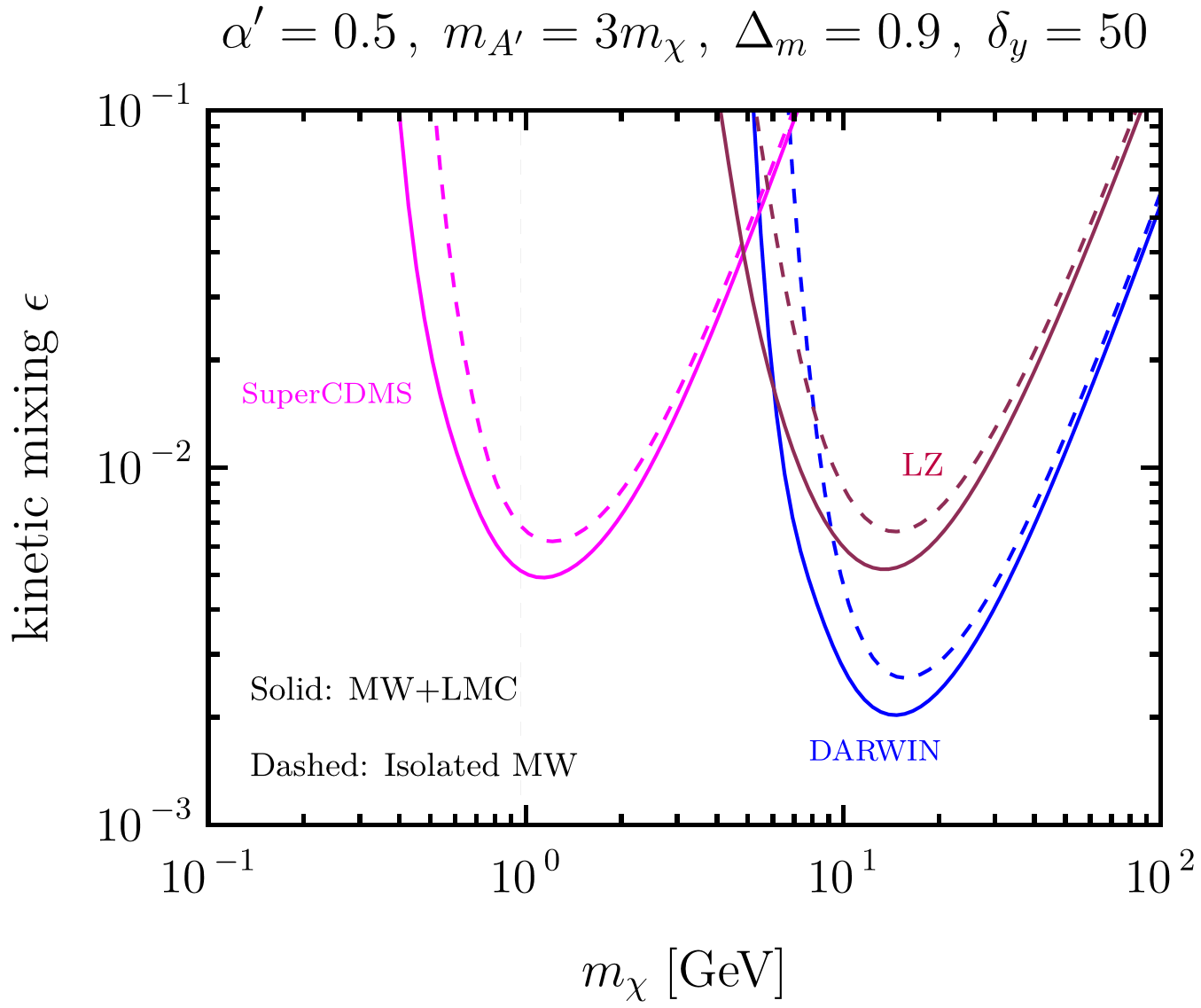}
    \caption{Comparison of exclusion limits and projected sensitivities for niDM from direct detection experiments using different local DM distributions. Solid lines are estimates obtained using the distributions for a Milk Way-like galaxy with a Large Magellanic Cloud-like satellite galaxy while dashed lines stand for an isolated Milk Way-like galaxy. Both distributions were computed in ref.~\cite{Reynoso-Cordova:2024xqz}. The isolated Milk Way estimations are in good agreement with the Standard Halo Model.}
    \label{fig:comparison-DD}
\end{figure}    

Using N-body simulations, ref.~\cite{Garavito-Camargo:2019kxw} concluded that the Large Magelanic Cloud (LMC) may have significant effects on the local DM density and velocity distribution. These changes were shown to have a significant impact for direct detection experiments~\cite{Smith-Orlik:2023kyl}. Very recently, using the Auriga cosmological simulations, ref.~\cite{Reynoso-Cordova:2024xqz} found that the lowest mass reach of direct detection experiments can be increased by around 10\% depending on the specific detector analyzed and the type of DM-nucleus interactions considered. Since the LMC influence is more pronounced for velocity-dependent operators, such as the effective operator $\mathcal{O}_8^N$ of niDM models, we investigate its impact on our direct detection computations.

In \cref{fig:comparison-DD}, we compare how the different assumptions for the DM distribution influence the direct detection exclusion and sensitivities for the niDM model. The LMC results are showed in comparison with those for an isolated Milky Way (MW). The latter was verified to agree with estimations using the Standard Halo Model with a local circular speed of 220 km/s. In particular, the SHM  assumes that the local DM velocity follows a Maxwell-Boltzmann distribution with a peak speed set to the local circular speed in the MW. The comparison shows an improvement of about 10\% for the lowest DM masses probed by direct detection when taking into account the LMC influence. This is relevant for niDM since these masses are not expected to be probed in the near-future by any other kind of experimental searches. 
    
}

{\section{Beam dump experiments}
\label{app:beam-dump}

In this appendix, we briefly discuss the experiments we consider: NuCal, CHARM, BEBC, and NA62 in the dump mode. We then compare our implementation with  ref.~\cite{Tsai:2019buq}, which has studied both dark photon and iDM models. A comparison between the constraints and sensitivities of these four different experiments for the niDM model can be seen in \cref{fig:allBeampDumps}.

BEBC is considered for the first time in the context of iDM, and it turns out that it provides constraints at a comparable level as NuCal (see \cref{fig:comparison-tsai}). The constraints it provides for small DM masses and lifetimes are similar to those obtained from NuCal, while for large masses and lifetimes, they are even stronger. Our estimates of the constraints imposed by NA62 show that they dominate over all other constraints coming from beam dump experiments, due a combination of a relatively large angular coverage and a high-energy proton beam.

\begin{figure}
     \centering
     \begin{subfigure}{0.42\textwidth}
        \centering
        \includegraphics[width=.99\linewidth]{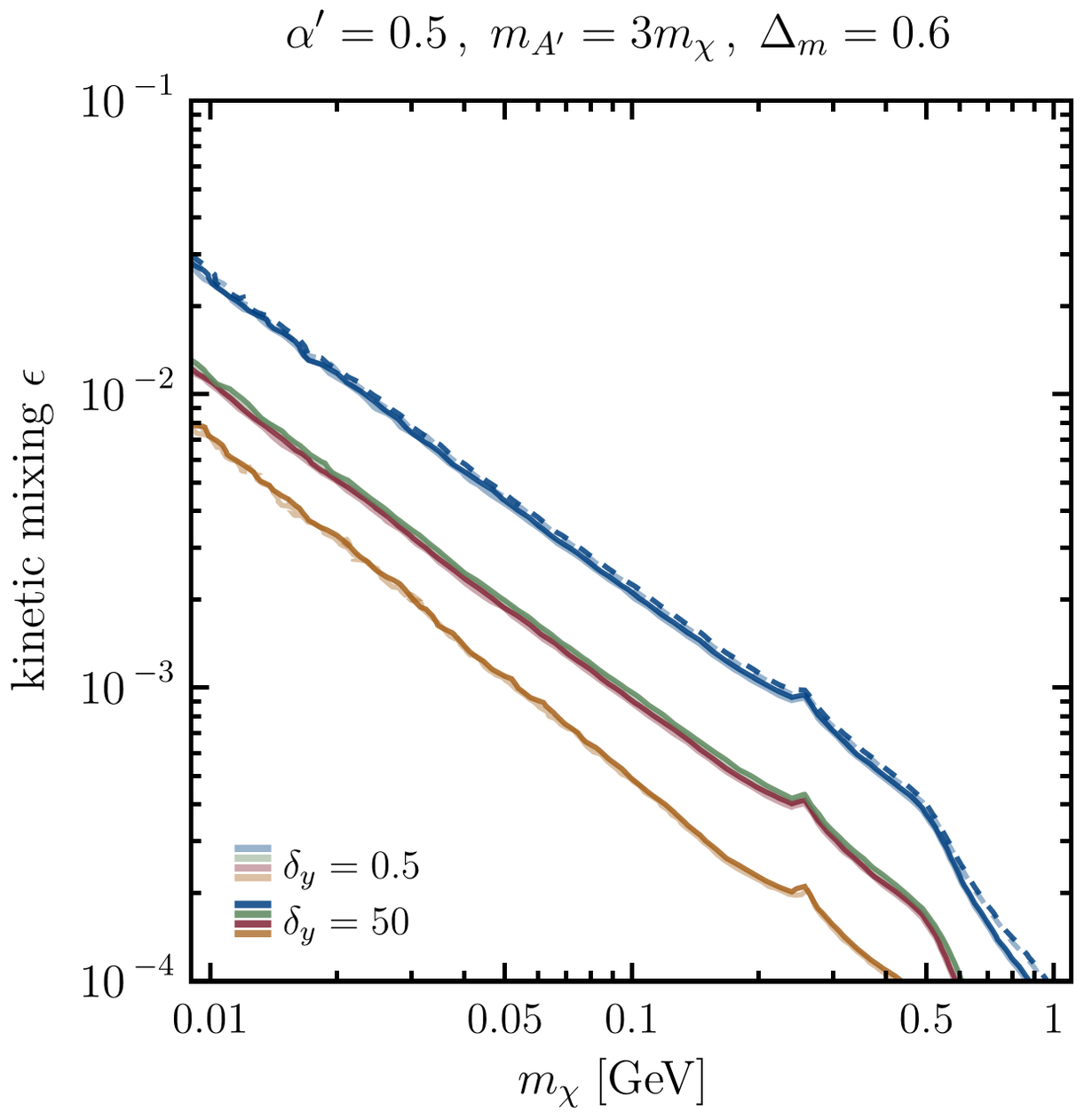}
     \end{subfigure}
     \hspace{0.05cm}
      \begin{subfigure}{0.42\textwidth}
        \centering
        \includegraphics[width=.99\linewidth]{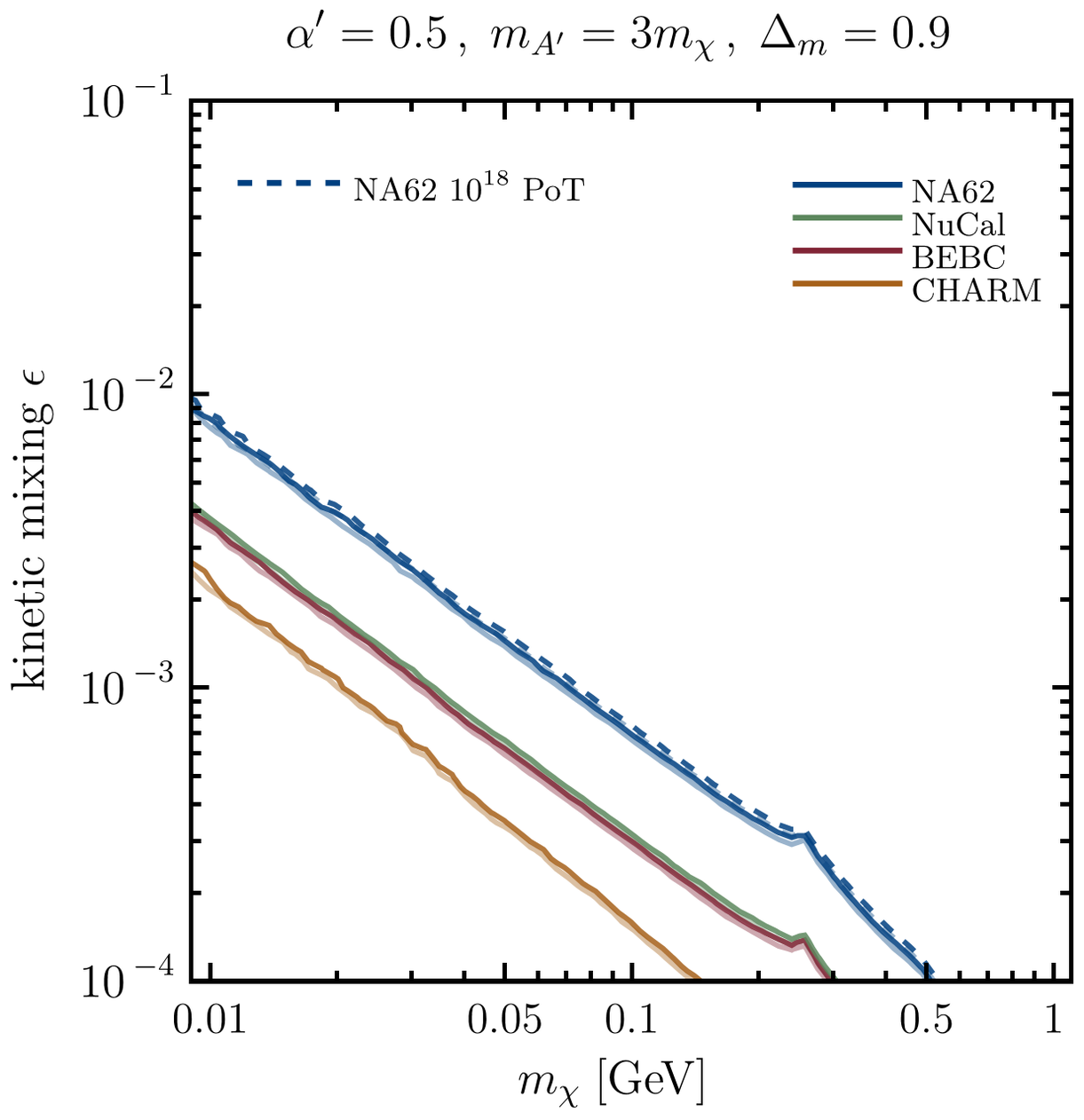}
     \end{subfigure}
        \caption{Current bounds (solid) and projected sensitivities (dashed) for the kinetic mixing $\epsilon$ from existing beam dump experiments. Each panel stands for a different combination of the relative mass splitting between dark sector fermions~$\Delta_m = (0.6,0.9)$. The opacity of the lines relates to different choices of dark left-right Yukawa asymmetry~$\delta_y = (0.5,50)$.}
        \label{fig:allBeampDumps}
\end{figure}

\subsection{NuCal}

NuCal~\cite{Blumlein:1990ay} was an experiment operated at Serpukhov's facility in 1989. A 70 GeV proton beam from the U70 accelerator with a total number of PoT of $N_{\text{PoT}} = 1.7\cdot 10^{18}$ was dumped on an iron target. The decay volume was a $2.6\times 2.6\times 23\text{ m}^{3}$ box located 64 m downstream from the target.  
For the event selection, we follow ref.~\cite{Tsai:2019buq} (and references therein): the pair of decay products (either $e^{+}e^{-}$ or $\mu^{+}\mu^{-}$) must point to the detector located at the end of the decay volume, and have total energy greater than $3\text{ GeV}$. The total reconstruction efficiency is assumed to be 70\%. 5 events have been observed, whereas the predicted SM background is at the level of 3.5 events implying that the 90\% CL exclusion limits correspond to $\approx 6$ events. Note that ref.~\cite{Tsai:2019buq} conservatively assumed a cylindrical decay volume (following the approximation made in ref~\cite{Blumlein:2013cua}).

\subsection{CHARM}
CHARM~\cite{CHARM:1983ayi,CHARM:1985anb,CHARM:1985nku} was a beam dump experiment operated at SPS in the 1980s. A proton beam with $N_{\text{PoT}}=2.4\cdot 10^{18}$ and $E_{p} = 400 \, \mathrm{GeV}$ collided with a copper target. The decay volume with dimensions $3\times3\times35\text{ m}^3$ was located 480 m downstream of the target; the transverse displacement of the center of the decay volume from the beam line was 5 m. 

For the selection of the decay products, we follow ref.~\cite{CHARM:1985nku}, which studied heavy neutral leptons and their decays $N\to \nu + ee/\mu\mu$, which are topologically similar to the decay processes of $\chi^{*}$. The decay products were required to be within the acceptance of the last layer of the calorimeter, located $\simeq 5$ m behind the end of the decay volume, and deposit at least 1 GeV of energy.

\subsection{BEBC}

BEBC was a bubble chamber experiment operated at SPS in the 1980s~\cite{BEBCWA66:1986err,WA66:1985mfx}. Unlike CHARM, its decay volume was on-axis, with dimensions $3.6\times 2.5\times 1.85\text{ m}^{3}$, located 404 m downstream of the target. The decay volume itself serves as a detector -- so there is no requirement for the trajectories of the decay products. The recent paper~\cite{Barouki:2022bkt} calculated the constraints from BEBC to heavy neutral leptons, considering, among the others, the decay channels $N\to ee\nu,\mu\mu\nu$ which are similar to the main decays of $\chi^\ast$ assuming a small mass difference $\Delta_m$. The requirement for the decay products is to pass the energy cut $E>1\text{ GeV}$, with an overall efficiency of 0.96. No events have been found for these modes. 

Compared to CHARM, the on-axis placement leads to a much larger flux of dark photons with large energies from bremsstrahlung. This, together with a smaller distance to the decay volume, significantly improves the yield of events in the domain of small lifetimes.

\subsection{NA62 in the dump mode}

NA62~\cite{NA62:2017rwk} is the experiment currently located at the ECN3 facility at CERN served by the SPS beam. Its primary goal is to study rare kaon decays. However, it may also work in the beam dump mode, with thick movable copper-iron collimators 
called TAXes serving as a target for the incoming 400 GeV proton beam. The effective decay volume~\cite{Jerhot:2022chi} is located 80 m downstream of the target; it has a cylindrical shape with a radius of 1 m, with a hole of $0.08$ cm radius accounting for the beam pipe; therefore. The detector system starts $\sim 81$ m behind the decay volume entrance and has a length of $\sim 64\text{ m}$. It also has an approximate cylindrical shape, with a radius of $1.15$ m, and the same hole accounting for the remnants of the incoming beam. In 2021, $1.4\cdot 10^{17}$ PoT have been collected~\cite{NA62:2023nhs} and a total of $\sim 10^{18}$ PoT are expected to be collected before LS3~\cite{Antel:2023hkf}.

We follow ref.~\cite{NA62:2023nhs} and impose the following requirements: the decay products must be within the acceptance of the detector up to the calorimeters located at its end; their momentum must be $>1\text{ GeV}$; the minimal transverse spatial displacement between the muons at the beginning of the straw tracker must be 2 cm; the transverse spatial separation between the electrons or hadrons at the entrance of electromagnetic calorimeter (ECAL) must be 20 cm, corresponding to a spatial separation of 95\% of the electromagnetic shower energy from the charged particles in liquid krypton.

In our analysis we have used the event selection similar to that of the search for dark photons. Unlike the latter, in events with niDM the full momentum of the decaying $\chi^{\ast}$ cannot be reconstructed since a part of it is carried by the invisible $\chi$. Therefore, in particular, the reconstructed 3-momentum of the lepton pair would not point directly to the target. This feature may be shared by a potential background. Detailed studies are required to analyse the impact parameter distribution of backgrounds and find the optimal transverse impact parameter cut for choosing the niDM events. However, we do not expect that the impact parameter would significantly affect the constraints imposed on large values of the coupling, which is the only parameter space relevant to our studies. Indeed, the upper bound is sensitive to the ratio $p_{\text{max}}/l_{\text{to decay volume}}$, where $p_{\text{max}}$ is the maximal momentum of the decaying particle and $l_{\text{to decay volume}}$ is the distance from the target to the beginning of the decay volume; other parameters defining the experiment enter the estimate logarithmically. The impact parameter cut would rather affect the events with low momentum of $\chi^{\ast}$, hence affecting the constraint slightly.

\subsection{Comparison with past works}

Let us now turn to the comparison of our analysis of the various constraints and sensitivities with ref.~\cite{Tsai:2019buq}. There are several notable differences in the modeling of the production of dark photons and the experimental setup. 

As for the dark photons, the first difference is that we include several additional dark photon production processes, such as the mixing with $\rho^{0}$ and the Drell-Yan process. The latter dominates the production of dark photons with masses $m_{A'} \gtrsim 2\text{ GeV}$ at SPS energies~\cite{SHiP:2020vbd}; the systematic uncertainties in the production cross section are within an acceptable 40\%. This channel is unimportant for dark photon searches at NA62, because the sensitivity does not extend to such large dark photon masses. However, this is no longer the case for iDM, and the Drell-Yan production channel has been found to be relevant for large DM masses. We do not include it in the comparison in this appendix, but we use it for analysis in the main text. 

Furthermore, we use a different proton form factor for proton bremsstrahlung, which acquires contributions from the $\rho$ and $\omega$ resonances~\cite{Feng:2017uoz} that have not been taken into account in~\cite{Tsai:2019buq}.\footnote{It is worth mentioning that there are large theoretical uncertainties in calculating the bremsstrahlung rate, so the results from the bremsstrahlung have to be taken with care.} Another difference concerns the number and distribution of neutral mesons $\pi^{0},\eta,\eta'$ used in the calculations. Ref.~\cite{Tsai:2019buq} approximated them from the distributions of charged pions taken from ref.~\cite{Bonesini:2001iz}. We instead follow the approach of ref.~\cite{Dobrich:2019dxc} when simulating the angle-energy distributions at all relevant facilities. 

\begin{figure}
    \centering
    \centering
     \begin{subfigure}{0.42\textwidth}
        \centering
        \includegraphics[width=.99\linewidth]{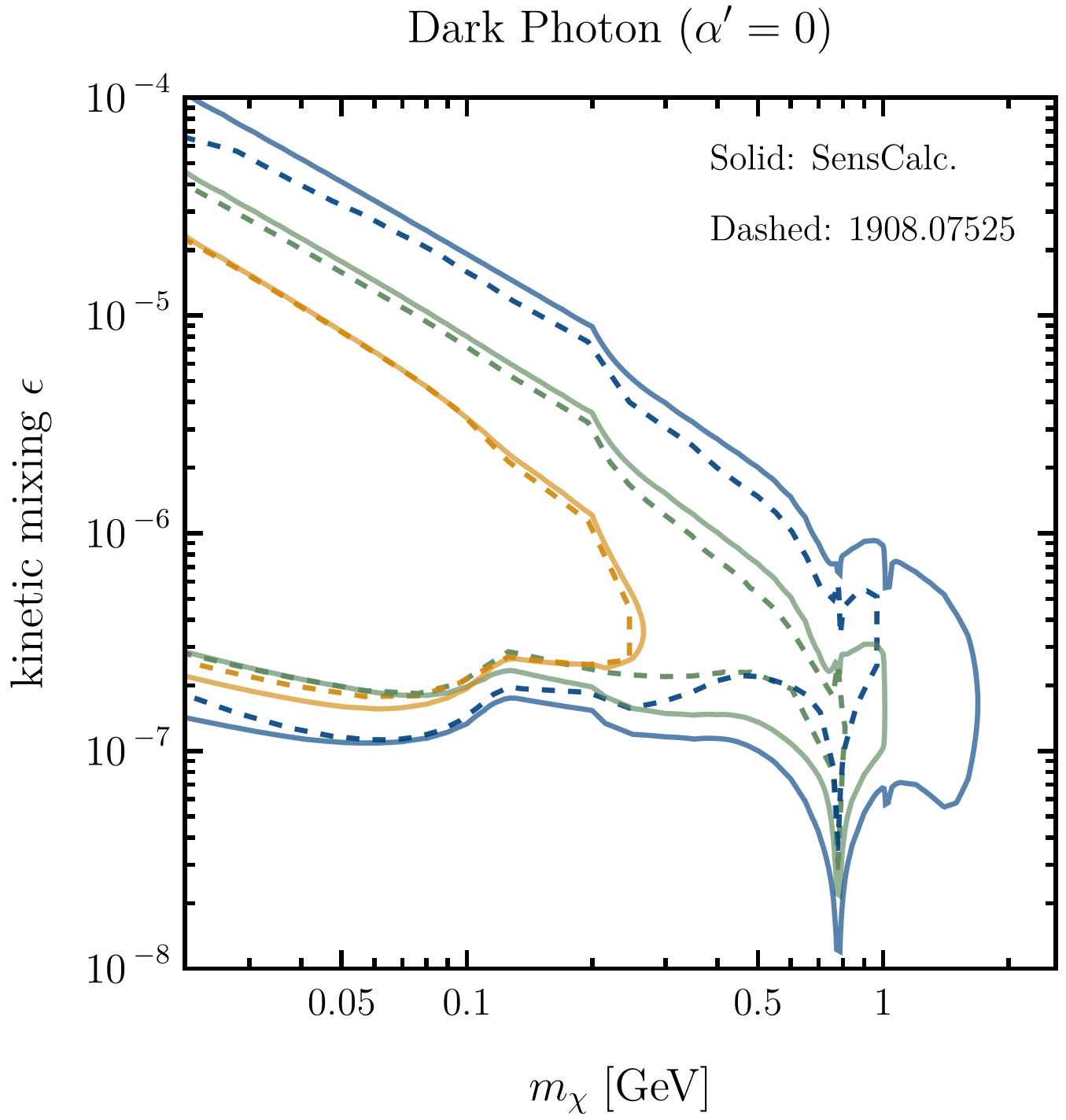}
     \end{subfigure}
     \hspace{0.05cm}
      \begin{subfigure}{0.42\textwidth}
        \centering
        \includegraphics[width=.99\linewidth]{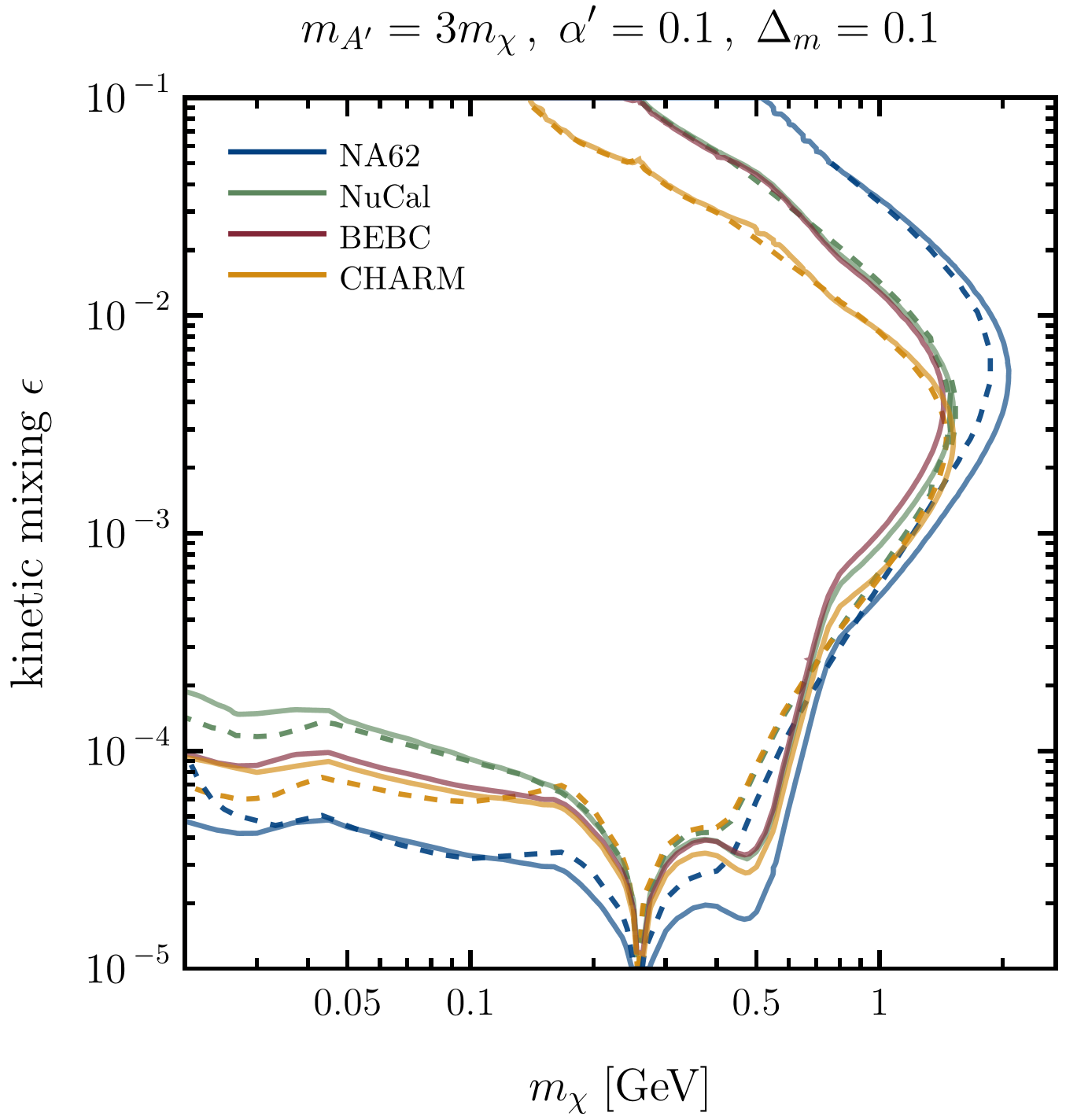}
     \end{subfigure}
    \caption{Comparison of exclusion limits and sensitivity projections for various beam dump experiments as obtained in this paper (solid lines) and in ref.~\cite{Tsai:2019buq} (dashed lines). Similarly to~\cite{Tsai:2019buq}, we assume that only leptonic final states are detectable. \textit{Left panel}: dark photons with vanishing invisible branching ratio. \textit{Right panel}: iDM coupled to dark photons, assuming $m_{A'} = 3 m_{\chi}$ and $\alpha' = 0.1$.}
    \label{fig:comparison-tsai}
\end{figure}

There are also some differences in the experimental setups. Regarding NA62, in ref.~\cite{Tsai:2019buq}, the decay products were required to have a transverse spatial separation of at least 10 cm at the entrance of the liquid krypton calorimeter, falsely assumed to be located immediately downward the decay volume. In reality, it is located $\simeq 50$ m downward, which significantly relaxes the impact of this cut. In addition, we include the magnetic field of the dipole magnet, which significantly affects the spatial separation of the decay products.

We present a comparison of the calculated exclusion bounds and sensitivity projections in \cref{fig:comparison-tsai}. For dark photons (left panel), the constraints we obtain agree in the parameter region where dark photons are produced dominantly by decays of mesons (up to the normalization of the meson fluxes), even despite the different description of the meson production, and start disagreeing for larger masses, where the dominant production channel is proton bremsstrahlung. This is a consequence of the different proton form factors. For iDM (right), we generally find a good agreement, with small differences in large masses.
}

{\section{Rescasting NA64 missing energy searches}
\label{app:NA64details}
The NA64 search conventionally targets dark photons that decay purely invisibly. In our set-up, on the other hand, the dark photon decays produce an excited state $\chi^\ast$, which may decay further within the detector and lead to a signal that is vetoed by the analysis. The number of events that pass the NA64 event selection for a given set of niDM parameters $\mathbf{x}_{\text{niDM}}=(m_{A'},\delta_y,\Delta_m,m_{\chi},\alpha',\epsilon)$ can therefore be written as:
\begin{equation} 
    N_{\text{niDM}}(\mathbf{x}_{\text{niDM}}) = N_{A'\to2\chi}(m_{A'}) - \epsilon^2 \left[ \sum_{i=\text{br},\text{an}}f_i(m_{A'}) g(\mathbf{a}_i,\mathbf{x}_{\text{niDM}}) \right] \, .
\end{equation}
Here the first term corresponds to the total number of events expected if the dark photon were to decay into a pair of stable invisible particles. 
In the second term, we have split the dark photon production into two different channels: production via electron-nucleus bremsstrahlung, $ e^- Z \to e^- Z A' $ ($i = \text{br}$), and via resonant annihilation of secondary positrons with atomic electrons, $e^+ e^- \to A'  $ ($i = \text{an}$). We denote the corresponding production rates by $f_i$, which are normalised such that
\begin{equation}
N_{A'\to2\chi}^{\text{90\% exc.}}(m_{A'}) = \epsilon^2_{\text{90\% exc.}}\left[f_{\text{br}}(m_{A'})+f_{\text{an}}(m_{A'})\right]
\end{equation}
using the $\epsilon_{\text{90\% exc.}}$ bounds given by the NA64 collaboration~\cite{Banerjee:2019pds,NA64:2023wbi}. The functions $g_i$ give the fraction of events vetoed by NA64 due to $\chi^\ast$ decays in the detector, such that $N_{\text{niDM}} = N_{A'\to2\chi}(m_{A'})$ for $g_i = 0$ and $N_{\text{niDM}} = 0$ for $g_i = 1$. These veto functions depend on the specific experimental conditions of NA64, collectively denoted by $\mathbf{a}_i$.

The veto functions $g_i$ decompose into two terms. The first describes decays of $\chi^\ast$ inside the ECAL, which decrease the total missing energy of the event and may lead to an event being vetoed if $E_{\text{missing}} < 50$~GeV. The second describes decays that would leave a non-zero energy deposit either in the VETO or within the HCAL, both of which cause an event to be vetoed. A general form for the veto function $g$ can be written as
\begin{align}
\label{eq:gFunctionForNA64}
g(\mathbf{a}_i,\mathbf{x}_{\text{niDM}}) = & \Omega_{i}\left(1-e^{-(L_{\text{ECAL}}-d_i^{\text{low}})/\ell_i^{\text{low}}}\right) \nonumber \\ & + \eta e^{-(L_{\text{HCAL}}-d_i^{\text{ave}}-l)/\ell_i^{\text{ave}}}\left(1-e^{-(L_{\text{HCAL}}+l)/\ell_i^{\text{ave}}}\right)  
\end{align}
where \begin{equation}
\ell_i^j = \frac{\tau_{\chi^{\ast}} \sqrt{(E_{\chi^\ast,i}^j)^2 - m_{\chi^\ast}^2} }{m_{\chi^{\ast}}} 
\end{equation} is the decay length of $\chi^{\ast}$, which we calculate separately for low-energy dark photons ($j = \text{low}$) and for all dark photons ($j = \text{ave}$). All parameters appearing in the veto functions $g$ are explained in detail in~\cref{tab:NA64parameters}. Finally, we assume that $E_{\chi^\ast,i}^j = E_{i}^j/2$, i.e., that the dark photon energy is equally shared between the two daughter particles. This is a reasonable approximation to adopt since dark photon energies are much larger than the masses of the final state particles.

\begin{table}
    \centering
    \caption{List of parameters used in the veto functions needed to reinterpret missing energy bounds from the NA64 experiment~\cite{Banerjee:2019pds,NA64:2023wbi}. The values presented in the first lines correspond to the channel $i=\text{bremsstrahlung}$, the only channel used for the fits, while those in the second lines to $i=\text{annihilation}$.}
    \begin{tabular}{llll} 
 \hline \hline\\[-4mm]
Description & Symbol & Estimated & Fitted \\
& & value & value    \\[1mm]
   \hline\\[-4mm]
 Fraction of produced dark photons & $\Omega_\text{br}$  & 20\% & 4\% \\ 
  with low energy ($50\,\mathrm{GeV} \leq E_{A'} \leq 75\,\mathrm{GeV}$) & $\Omega_\text{an}$ & 50\% &  -- \\[2mm]
  Average energy of low energy dark & $E_\text{br}^\text{low}$ & 64 GeV & 74 GeV \\
  photons producing a signal & $E_\text{an}^\text{low}$ & Eq.~\eqref{eq:Eave} & -- \\[2mm]
    Low energy dark photon production & $d_\text{br}^{\text{low}}$ & $3X_0$ & $22 X_0$  \\
 position plus containment length\tablefootnote{This parameter defines an effective ECAL length for low energy dark photon events.} & $d_{\text{an}}^{\text{low}}$ & $d_{\text{br}}^{\text{low}} + 2X_0$ & --    \\[1mm]
   \hline\\[-4mm]
 Average energy of dark photons & $E_\text{br}^\text{ave}$ & 85 GeV & 78 GeV \\
producing a signal & $E_\text{an}^\text{ave}$ & Eq.~\eqref{eq:Eave} & -- \\[2mm]
   Average energy dark photon & $d_\text{br}^{\text{ave}}$ & $X_0$ & $11X_0$ \\
   production position & $d_\text{an}^{\text{ave}}$ & $d_{\text{br}}^{\text{ave}} + 2X_0$ & -- 
   \\[1mm]
   \hline\\[-4mm]
 $A'$ veto efficiency for events & \multirow{2}{*}{$\eta$}  & \multirow{2}{*}{100\%}  &\multirow{2}{*}{100\%} \\ 
  leaving hits at VETO + HCAL & &  &  \\[2mm]
ECAL length & $L_{\text{ECAL}}$ & 1 m ($40 X_0$) & -- \\[2mm]
 HCAL length & $L_{\text{HCAL}}$ & 6.5 m &  -- \\[2mm]
 Leakage length of showers produced at & \multirow{2}{*}{$l$}& \multirow{2}{*}{$5X_0$} & \multirow{2}{*}{$7X_0$} \\
  ECAL leacking energy to VETO + HCAL &  &   &  \\[1mm] \\
 \hline
 \hline
\end{tabular}
    \label{tab:NA64parameters}
\end{table}

With a physical model for the veto functions in NA64, we can compare our predictions to those for iDM available in the literature. Doing so allows for a validation of our parametrisation and for a determination of the various parameters that we have introduced. For the purpose of this comparison, we set $f_{\text{an}} = 0$, since this channel has not yet been considered for the iDM scenario~\cite{Mongillo:2023hbs,Abdullahi:2023tyk}. As a first step, we estimate the parameters $\mathbf{a}_{\text{br}}$ using the available information on the experimental conditions~\cite{Gninenko:2767779} and previous studies~\cite{Gninenko:2017yus}, see the third column in \cref{tab:NA64parameters}.  For the iDM case ($\delta_y = 0$), our results agree with ref.~\cite{Abdullahi:2023tyk} within 14\% (with an averaged error of about 6\%), see \cref{fig:comparison-NA64}. To further improve the agreement, we vary the parameters $\mathbf{a}_{\text{br}}$ around their estimated value and find the best fit of our function to the results from ref.~\cite{Abdullahi:2023tyk}, see the fourth column in \cref{tab:NA64parameters}. The fit thus obtained matches both predictions of fig.~6 from ref.~\cite{Abdullahi:2023tyk} with a maximal discrepancy of less than 8\% (with an averaged error of less than 2\%), see once again \cref{fig:comparison-NA64}. Contrary to our initial expectation, we find $\Omega_{\text{br}} \ll 1$, which suggests that decays of $\chi^\ast$ particles produced from low-energy bremsstrahlung dark photons, $E_{A'} \in (50,75)$~GeV, are irrelevant for the constraints. We use the fitted parameters to give our final estimates for the bremsstrahlung produced signals (which is the most relevant production channel over nearly all the probed parameter space) at NA64.
 
Since previous studies have not considered dark photon signals produced by positron annihilation~\cite{Andreev:2021fzd}, we cannot compare our estimated values for $\mathbf{a}_{\text{an}}$ to the literature as we did for the bremsstrahlung channel. We can, however, make use of fig~1 of ref.~\cite{Andreev:2021fzd} to estimate the average energy of dark photons produced via positrons which can generate a missing energy event:\footnote{Since the average energy $E_{\text{an}}^{\text{ave}}$ is generally low for the region where the annihilation channel is relevant, we approximate $E_{\text{an}}^{\text{low}}=E_{\text{an}}^{\text{ave}}$.}
\begin{equation}
\label{eq:Eave}
     E_{\text{an}}^{\text{ave}}(m_{A'}) = \begin{cases}
		50\,\mathrm{GeV}, & m_{A'} < 225 \, \mathrm{MeV}\\
            430 m_{A'} - 47 \, \mathrm{GeV}, & m_{A'} \geq  225 \, \mathrm{MeV}.
		 \end{cases} 
 \end{equation}
 We then combine both channels in order to recast NA64 bounds for niDM with the most current available data from NA64~\cite{NA64:2023wbi}. 

 \begin{figure}
    \centering
    \includegraphics[width=0.5\textwidth]{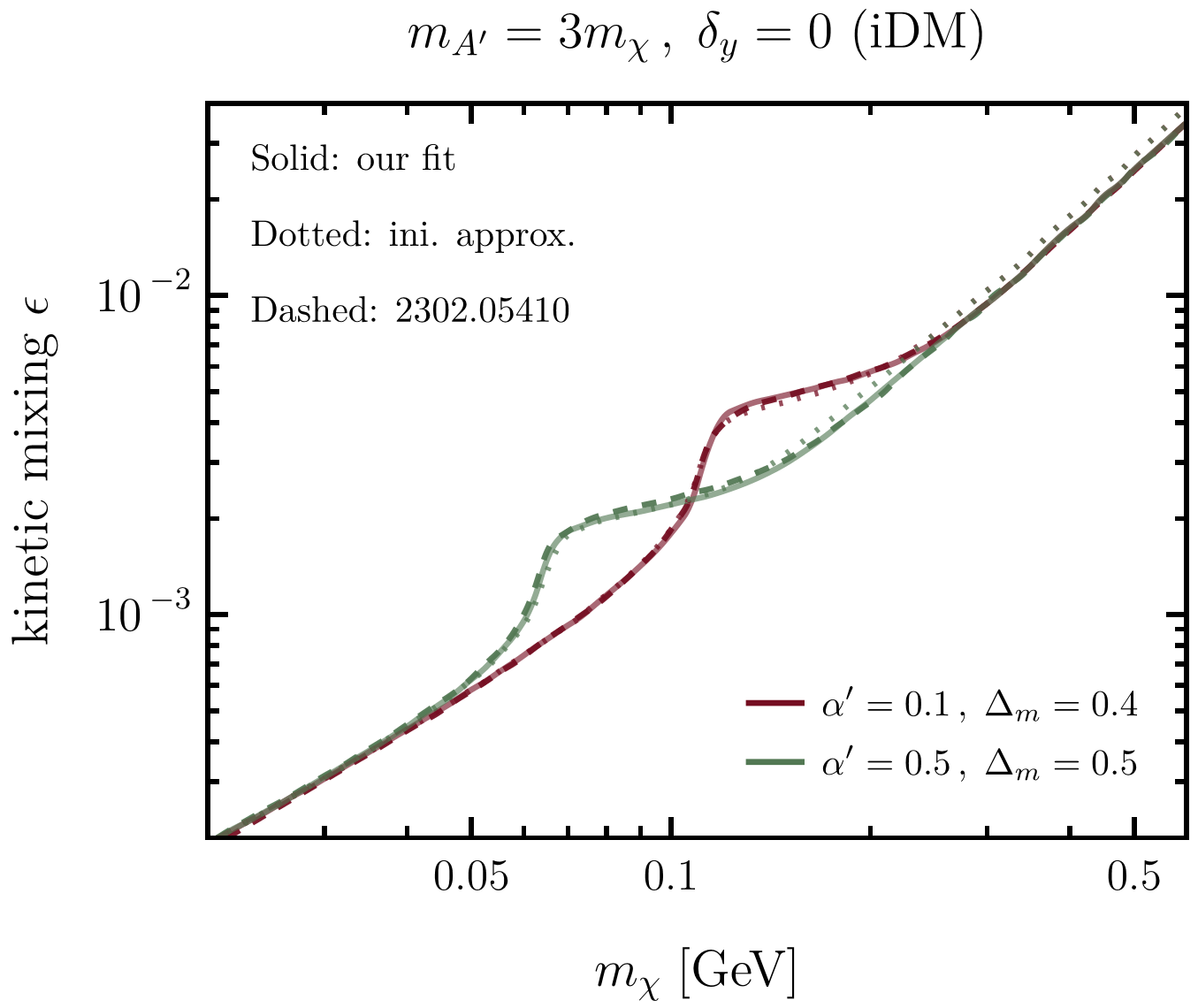}
    \caption{Comparison of exclusion limits on iDM for the NA64 experiment with $2.84\times 10^{11}$ EoT as obtained in this paper (dotted and solid lines) and in ref.~\cite{Abdullahi:2023tyk} (dashed lines). The dotted curves are obtained by estimating the parameters that enter the veto function from experimental data, the solid curves are obtained by fitting these parameters to reproduce  the results from ref.~\cite{Abdullahi:2023tyk} as closely as possible. Dark photons produced from secondary positron were not considered in the comparison since they were neglected in the original NA64 analysis~\cite{Banerjee:2019pds} as well as in the reinterpretation done in ref.~\cite{Abdullahi:2023tyk}.}
    \label{fig:comparison-NA64}
\end{figure}
}

{\section{Set-up for electron-positron colliders}
\label{app:colliders}
In this appendix, we briefly introduce the electron-positron collider experiments that we consider in this work, BaBar and Belle~II, their basic properties relevant for our simulations and the single-photon selection cuts required for the event analysis. We also present in \cref{fig:comparison-BABARbelleII} the comparison of our simulation in the iDM limit $\delta_y=0$ with results from ref.~\cite{Duerr:2019dmv}.

\subsection{BaBar}

The BaBar collaboration searched for missing energy events in a dataset of 53~fb$^{-1}$~\cite{BaBar:2017tiz} produced by the PEP-II assymetric $e^+e^-$ collider at the SLAC National Accelerator
Laboratory with center-of-mass~(CM) energy of $\sqrt{s} = 10.58$~GeV, the mass of the $\Upsilon(4S)$ resonance.\footnote{Although, most of the data from the BaBar detector has been collected at the $\Upsilon(4S)$~\cite{BaBar:2013agn}, the data used at the analysis was mainly collected at the $\Upsilon(3S)$ resonance and non-negligeble amounts of data were also taken at the resonances $\Upsilon(2S)$ and “off-resonances"~\cite{BaBar:2017tiz}. However, within the precision required in this work, the approximation  $\sqrt{s} = 10.58$~GeV is sufficient.} The positron beam of 3.1 GeV and the electron beam of 9.0 GeV collide head-on in PEP-II implying a velocity of the CM along the detector’s magnetic
field axis of $\beta_z \approx 0.49$~\cite{BaBar:2001yhh}.

To analyse BaBar's single-photon searches, we follow the SM charged final state particles selection cuts from table~2 of ref.~\cite{Duerr:2019dmv} which aim to reproduce the more complex multivariate BaBar analysis. We also use their photon selection criteria, namely $E_{\gamma}^{\text{CM}} > 2.0$~GeV and $ 32.5^\circ < \theta_{\gamma}^{\text{CM}} < 99^\circ$ where the angle $\theta$ is with respect to the positive $z$-axis. 

\subsection{Belle II}

The potential of Belle II to search for invisibly decaying dark photons via single-photon searches was analysed in ref.~\cite{Belle-II:2018jsg}. Their estimates are given for an integrated luminosity of 20~fb$^{-1}$ generated by the SuperKEKB asymetric $e^+e^-$ collider at the High Energy Accelerator Research Organisation with a CM energy of  $m_{\Upsilon(4S)}$. The positron beam of 4 GeV and the electron beam of 7 GeV meet with an angle of 83 mrad, where the bisector of this angle goes along the detector's magnetic field axis, implying a CM velocity of $\Vec{\beta} \approx (0,0.04,0.27)$~\cite{Belle-II:2018jsg}.

Once more, we use the selection criteria provided by ref.~\cite{Duerr:2019dmv}, which provides the necessary cuts for charged particles in table 1. The photon selection is more involved than for BaBar and depends on the value of the dark photon mass. We adopt the detailed photon selection requirements given in eqs. (3.1-3.4) of ref.~\cite{Duerr:2019dmv}.

 \begin{figure}[t]
     \centering
     \begin{subfigure}{0.42\textwidth}
        \centering
        \includegraphics[width=.99\linewidth]{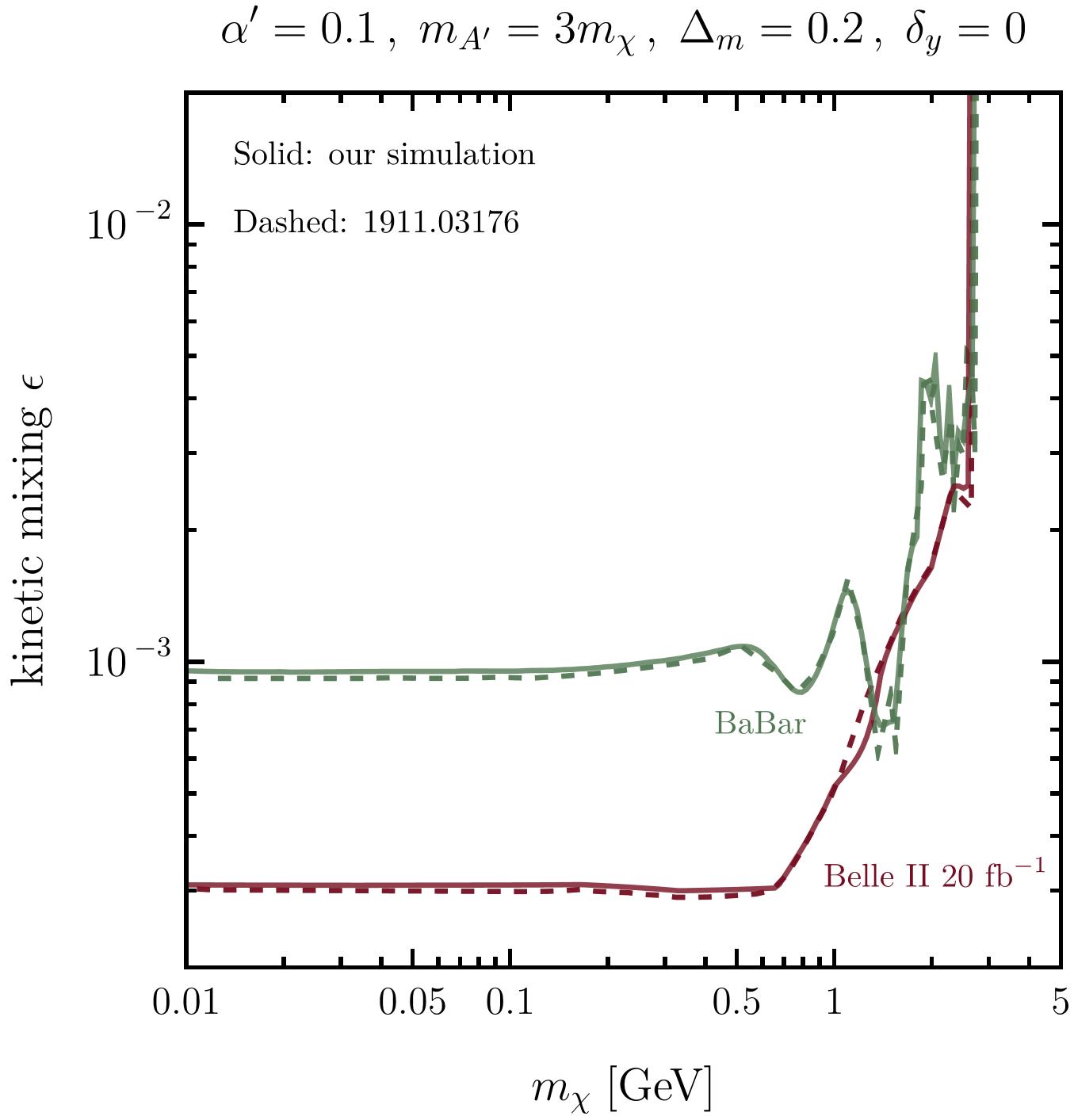}
     \end{subfigure}
     \hspace{0.05cm}
      \begin{subfigure}{0.42\textwidth}
        \centering
        \includegraphics[width=.99\linewidth]{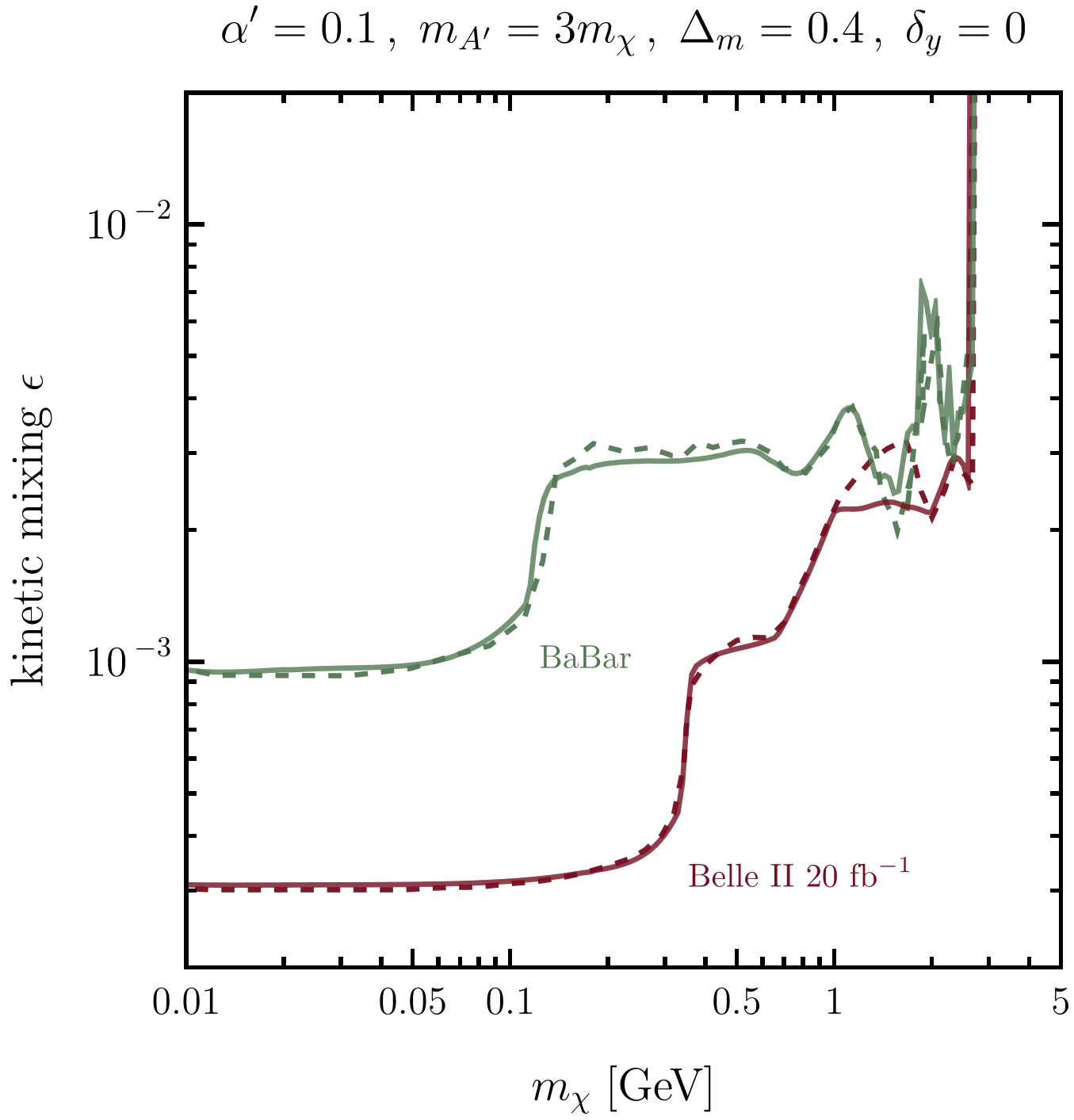}
     \end{subfigure}
    \caption{Comparison of BaBar exclusion limits and Belle~II projected sensitivities on iDM from mono-photon searches~\cite{BaBar:2017tiz,Belle-II:2018jsg} as obtained in this paper (solid lines) and in ref.~\cite{Duerr:2019dmv} (dashed lines). }
    \label{fig:comparison-BABARbelleII}
\end{figure}
}

\bibliographystyle{JHEP}
\bibliography{biblio.bib}

\end{document}